\documentstyle[agums,psfig]{article} 	    
\lefthead{OUILLON and SORNETTE}
\righthead{Magnitude-Dependent Omori Law}	
       \authoraddr{ Guy Ouillon, Department of Earth and Space Sciences and Institute of
Geophysics and Planetary Physics, University of California, Los Angeles,
California and Laboratoire de Physique de la Mati\`{e}re Condens\'{e}e,
CNRS UMR 6622 and Universit\'{e} de Nice-Sophia Antipolis, Parc Valrose,
06108 Nice, France  (e-mail: ouillon@aol.com)}
\authoraddr{Didier Sornette,
Department of Earth and Space Sciences and Institute of
Geophysics and Planetary Physics, University of California, Los Angeles,
California and Laboratoire de Physique de la Mati\`{e}re Condens\'{e}e,
CNRS UMR 6622 and Universit\'{e} de Nice-Sophia Antipolis, Parc Valrose,
06108 Nice, France (e-mail: sornette@moho.ess.ucla.edu)}

\input{update.tex}
\begin{document}
\title{Magnitude-Dependent Omori Law: Empirical Study and Theory}
\author{G. Ouillon and D. Sornette}
\affil{Department of Earth and Space Sciences and Institute of
Geophysics and Planetary Physics, University of California, Los Angeles,
      California 90095-1567 and
    Laboratoire de Physique de la Mati\`{e}re Condens\'{e}e, CNRS UMR 6622
Universit\'{e} de Nice-Sophia Antipolis, Parc Valrose, 06108 Nice, France}

\newcommand{\be}{\begin{equation}}
\newcommand{\ee}{\end{equation}}
\newcommand{\ba}{\begin{eqnarray}}
\newcommand{\ea}{\end{eqnarray}}
\newenvironment{technical}{\begin{quotation}\small}{\end{quotation}}
\renewcommand{\thefootnote}{\arabic{footnote}}
\newtheorem{definition}{Hypothesis}


\begin{abstract}

We propose a new physically-based ``multifractal stress activation''
model of earthquake interaction and triggering
based on two simple ingredients: (i) a seismic rupture results from activated processes
giving an exponential dependence on the local stress; (ii) the stress relaxation has
a long memory. The combination of these two effects predicts in a rather general way
that seismic decay rates after mainshocks follow the 
Omori law $\sim 1/t^p$ with exponents $p$ linearly increasing with the 
magnitude $M_L$ of the mainshock and the inverse temperature. 
We carefully test the prediction on the magnitude dependence of $p$
by a detailed analysis of earthquake
sequences in the Southern California Earthquake catalog. We find
power law relaxations 
of seismic sequences triggered by mainshocks with exponents
$p$ increasing with the mainshock magnitude by approximately $0.1-0.15$
for each magnitude unit increase, from $p(M_L=3) \approx 0.6$ to $p(M_L=7) \approx 1.1$,
in good agreement with the prediction of the multifractal model. 
The results are robust with respect 
to different time intervals, magnitude ranges and
declustering methods. When applied to synthetic catalogs generated
by the ETAS (Epidemic-Type Aftershock Sequence) model constituting 
a strong null hypothesis with built-in magnitude-independent $p$-values,
our procedure recovers the correct magnitude-independent $p$-values.
Our analysis thus suggests that a new important fact of 
seismicity has been unearthed. We discuss alternative interpretations of the data and
describe other predictions of the model.
\end{abstract}

\begin{article}

\section{Introduction}

There are now many evidences that the space-time organization of earthquakes
is consistent with the idea that a single
physical triggering mechanism is responsible for the occurrence of
aftershocks, mainshocks, foreshocks, and multiplets, leading to the more encompassing concept
of earthquake triggering (see for instance
[{\it Lin and Stein}, 2004; {\it Feltzer et al.}, 2004; {\it Murru et al.}, 2004;
{\it Huc and Main}, 2003; {\it Marsan}, 2003; {\it Helmstetter and Sornette}, 2003;
{\it Papazachos et al.}, 2000]). 

Earthquake triggering has been modeled using a variety of approaches including,
static stress transfer calculations [{\it King et al.}, 1994; {\it Stein}, 2003],
dynamic triggering processes [{\it Voisin}, 2002; {\it Perfettini et al.}, 2003],
as well as more phenomenological epidemic-type aftershock sequence (ETAS) models
[{\it Kagan and Knopoff}, 1981; 1987; {\it Ogata}, 1988] which are based 
on ``mutually self-excited point
processes'' introduced by {\it Hawkes} [1971] (see [{\it Helmstetter and Sornette}, 2002a]
for properties of the ETAS model and a review of the literature).
The later modeling approach in particular has been able to rationalize most of the 
phenomenological statistical properties of earthquake catalogs, such as 
the larger proportion than normal of large versus small foreshocks,
the power law acceleration of stacked seismicity rate as a function of
time to the mainshock, the spatial migration of foreshocks toward 
the mainshock when averaging over many sequences, the independence of 
foreshocks precursory properties as a function of the
mainshock size, the existence of correlations in seismicity 
over surprisingly large length scales [{\it Helmstetter and Sornette}, 2003a],
B{\aa}th's law [{\it Helmstetter and Sornette}, 2003b] and so on.

Its appeals due the simplicity of its premise, its power of explanation
of a large set of empirical observations
and the relative ease with which it can be implemented for 
rigorous statistical tests has made the class of ETAS models
the natural null hypothesis against which to test any other model of seismicity.
The class of ETAS models suffers however from a major drawback, that is, the lack
of a clear physical basis. The ETAS model is a statistical phenomenological construction which
postulates a dependence of the present 
seismic rate on past seismic rates propagated forward 
in time via a bare Omori propagator. 

One would thus like models in which
seismic rates derive from the interaction between and transfer of physical fields, 
such as stress, strain rates
and fluid flows. For this purpose, we present here a 
generalization which has a sound physical basis,
as it relies on two fundamental 
physical ingredients: rupture activation and stress transfer.
Rupture activation is described by the generic thermal activation processes
at the origin of the state-and-velocity dependent friction laws, 
stress corrosion effects, and so on. In the description of stress transfer,
we take into account the 
long-time memory effects in the relaxation of 
stress fields due to the visco-elasto-plastic rheology of the crust and
upper mantle. Using this model, we predict a new 
phenomenon, the dependence of the exponent $p$ of the Omori law
on the mainshock magnitude. This prediction is tested successfully by 
a careful analysis of the Southern California earthquake catalog.
For this and for other reasons that
will become clear below, we
coin our model the ``multifractal stress activation'' model. 

The organization of this paper is the following. In section 2, we first discuss
the fundamental physical ingredients of the ``multifractal stress activation'' model.
Section 3 gives its detailed definition.
Section 4 shows by analytical calculations that the Omori law $\sim 1/t^p$ derives naturally
from the model, but with an exponent $p$ which is an increasing linear function of the mainshock
magnitude. This surprising prediction results from the interplay between the exponential
activation process and the long-time memory of stress relaxation processes. Section 5 tests
this prediction using different declustering techniques to identify mainshocks at magnitudes
ranging from $1.5$ to $7.5$ in the Southern California catalog and to stack their corresponding
aftershock sequences. We find a remarkable agreement with $p(M)$ increasing from 
approximately $p=0.6$ for $M=3$ to $p=1.1$ for $M=7$. We also present tests on 
synthetic catalogs generated with the ETAS model.
Section 6 presents other observations that can be reinterpreted
within the framework of our model and discuss other predictions.

\section{Fundamentals of the ``multifractal stress activation'' model}

\subsection{Earthquakes as thermally activated processes \label{ss1}}

We model seismic activity as the occurence of
frictional sliding events and/or fault ruptures that are thermally activated processes 
facilitated by the applied stress field. The relevance of thermal activation is
made clear when examining the underlying physical processes of the various
proposed models of earthquakes, which we now briefly review.

\begin{itemize}
\item Thermal activation is known to control
creep rupture (also called static fatigue) (see for instance
the review in [{\it Scholz}, 2002] for the application to rock mechanics and
to earthquakes and 
[{\it Ciliberto et al.}, 2001; {\it Politi et al.}, 2002; {\it Saichev and Sornette}, 2003]
for recent experimental and theoretical developments).

\item The use of the Eyring rheology and other thermally-dependent
friction laws are of
standard use for describing creep failure in a variety of compounds 
[{\it Liu and Ross}, 1996] as well as
material interfaces [{\it Vulliet}, 2000].
These laws consist in adapting, at the microscopic level,
the theory of reaction rates
describing processes activated by crossing potential barriers.

\item Stress corrosion occurs in the presence of pre-existing cracks
in quartz, quartz rocks, calcite rocks, basaltic rocks, granitic rocks and many
other geological materials [{\it Atkinson et al.}, 1981; {\it Atkinson}, 1984]
by the mechanism of hydrolytic
weakening [{\it Griggs et al.}, 1957; {\it Griggs and Handin}, 1960] 
which is also thermally activated (see [{\it Sornette}, 1999] for a review
and references therein).

\item The Ruina-Dieterich state-and-velocity dependent friction law [{\it Dieterich}, 1979;
{\it Ruina}, 1983; {\it Scholz}, 1998]
results physically from creep of the surface contact and a
consequent increase in real contact area with time of contact
[{\it Scholz and Engelder}, 1976; {\it Wang and Scholz}, 1994].
The logarithmic form $\ln V$ of the velocity dependence of the friction
coefficient in the Ruina-Dieterich law is usually assumed to derive from
an Arrhenius activated rate process describing creep at asperity contacts
[{\it Stesky}, 1977; {\it Chester and Higgs}, 1992; 
{\it Chester}, 1994; {\it Heslot et al.}, 1994;
{\it Brechet and Estrin}, 1994; {\it Baumberger}, 1997; {\it Sleep}, 1997; 
{\it Persson}, 1998; {\it Baumberger et
al.}, 1999; {\it Lapusta et al.}, 2000; {\it Nakatani}, 2001].

\end{itemize}

\subsection{Long-time memory effects in the relaxation of stress fields \label{ss2}}

A recent re-construction of the regional strain rate map in California
from GPS recordings 
over a dense network shows that the largest strain rates are controlled by 
past large earthquakes and are found in
the regions where aftershock activity is still noticeable [{\it Jackson et al.}, 1997].
This suggests the relevance of a stress-controlled earthquake activation.
Post-seismic slip and strain rate relaxation following large earthquakes
have been modeled by visco-elastic flows, which govern 
the evolution of the stress field and thus the loading and
unloading processes of major earthquake generating
faults [{\it Deng et al.}, 1998]. The simplest models 
assume linear visco-elastic rheologies, which lead to exponential strain relaxation.
These models in general account for the short-term relaxation processes
over a time scale ranging from a few months to one year
[{\it Pollitz et al.}, 2000]. Over long time scales, it is necessary to 
take into account the presence and geometry of lower crustal and mantle shear zones,
which lead to more complex and slower decaying relaxation rates 
[{\it Kenner and Segall}, 1999]. 
To adequately model long-term postseismic relaxation (i.e. that
occurring years to decades after an earthquake) for instance
after the Landers and Hector Mine earthquakes, it was found necessary to add other
slower deformation processes [{\it Freed and Lin}, 2000; 2001].
Evidence of large post-seismic relaxation times have been found within the crust in
the case of slow-rate intracontinental events [{\it Calais et al.}, 2002].
Long-term stress relaxation process are also found in empirical laws of 
creep experiments applied to model the brittle creeping fault zone,  which 
can account for both the time evolution of
afterslip, as measured from geodesy, and of aftershocks decay. 
In this framework, aftershock sequences and deep afterslip, as
constrained from geodetic measurements, follow the same temporal
evolution [{\it Perfettini and Avouac}, 2004].
Generally, slower-than-exponential relaxation is found
in disordered materials which can be characterized by nonlinear
rheologies [{\it Klinger}, 1988; {\it Chung and Stevens}, 1991; {\it Phillips}, 1996].

\section{Formulation of the ``multifractal stress activation'' model}

\subsection{General case}

Let us denote by $\lambda({\vec r}, t)$ the intensity (or average
conditional seismicity rate) at position $\vec r$ and time $t$. 
Putting together the two physical ingredients of 
rupture activation and stress transfer discussed above, we formulate the following model.
The thermal rupture activation process is expressed as
\be
\lambda({\vec r}, t) \sim \exp \left[-\beta E ({\vec r}, t) \right]~,
\label{mvmmcs}
\ee
where $\beta$ is the inverse temperature (specifically $\beta =1/kT$
where $k$ is the Boltzmann constant) and the 
energy barrier $E({\vec r}, t)$ for rupture can be written as the sum of a contribution
$E_0({\vec r})$ 
characterizing the material and of a term linearly decreasing with the locally
applied stress $\Sigma({\vec r}, t)$ [{\it Zhurkov}, 1965]: 
\be
E({\vec r}, t) = E_0({\vec r}) - V \Sigma({\vec r}, t)~.
\label{gmjmwels}
\ee
$V$ is a constant which has the dimension of a volume and $\Sigma({\vec r}, t)$
is the total stress at position ${\vec r}$ and time $t$. The decrease
of the energy barrier $E({\vec r}, t)$ as a function of 
the applied stress $ \Sigma({\vec r}, t)$ in (\ref{gmjmwels}) embodies 
the various physical processes of stress corrosion, state-and-velocity
dependent friction and mechano-chemical effects, aiding rupture activation under stress.
The stress $\Sigma({\vec r}, t)$
results itself from all past events according to
\be
 \Sigma({\vec r}, t) = \Sigma_{\rm far~field}({\vec r}, t) + 
 \int_{-\infty}^t  \int dN[d{\vec r}~' \times d\tau]
\Delta \sigma({\vec r}~', \tau)  g({\vec r} - {\vec r}~', t-\tau)~.
\label{mgmmwsls}
\ee
Putting all this together yields
\be
\lambda({\vec r}, t) = \lambda_{\rm tec} ({\vec r}, t)  
\exp \left[\beta \int_{-\infty}^t  \int dN[d{\vec r}~' \times d\tau]
\Delta \sigma({\vec r}~', \tau) 
g({\vec r} - {\vec r}~', t-\tau) \right]~.
\label{fund}
\ee
In this expression, the term $\beta$ now incorporates the volume term $V$
and the inverse temperature ($\beta = V/kT$), so that $\beta$ has now the dimension of the inverse
of a stress. The double integral
gives the stress at position $\vec r$ and time $t$ as the sum 
of the stress load contributions over all
past earthquakes at earlier times $\tau<t$ and positions ${\vec r}~'$. A given
past event at ${\vec r}~'$ and time $\tau$ contributes to the stress at
$\vec r$ and time $t$ by its stress drop amplitude $\Delta \sigma({\vec r}~', \tau)$
which is transfered in space and time via the stress kernel (or Green function)
$g({\vec r} - {\vec r}', t-\tau)$, taking into account both time relaxation
and spatial geometrical decay. The kernel $g({\vec r} - {\vec r}~', t-\tau)$
describes the combined effects
of all stress relaxation processes in space and time, that determine
the stress field in the seismogenic layer, as discussed in section \ref{ss2}.
The term $dN[d{\vec r}~' \times d\tau]$ is the number of 
events in the volume $d{\vec r}~'$ that occurred between $\tau$ and $\tau+d\tau$.
Finally, $\lambda_{\rm tec} ({\vec r}, t)$ is the 
spontaneous seismicity rate in absence of stress triggering by other earthquakes
and accounts for the tectonic loading (far field stress), which may in general be non-homogeneous
and space and perhaps depends on time.

It is convenient to discretize space in cells and rewrite (\ref{fund}) as
\be
\lambda_i(t) = \lambda_{\rm tec}(t)  
\exp \left[ \beta \sum_j \int_{-\infty}^t  d\tau~
\Delta \sigma_j(\tau) ~ g_{ij}(t-\tau) \right]~,
\label{funddis}
\ee
where $\lambda_i(t)$ is the average conditional seismic rate in cell $i$ at time $t$,
$\Delta \sigma_j(\tau)$ is the stress drop in cell $j$ that occurred at time $\tau$
due to an earthquake and $g_{ij}(t-\tau)$ measures the fraction of the stress
drop that occurred at time $\tau$ in cell $j$ which is transfered to cell $j$ at time $t$.

Our model (\ref{fund},\ref{funddis}) is reminiscent of the stress release model, but
there are several important differences that are worth noting.
The single cell stress release
model was introduced by Vere-Jones [1978] as a stochastic implementation of 
Reid's theory of elastic rebound theory [{\it Reid}, 1910]. The 
generalization to account for long-range elastic stress transfer
was done in [{\it Zheng and Vere-Jones}, 1991] and in [{\it Liu et al.}, 1998;
{\it Shi et al.}, 1998] (see [{\it Bebbington and Harte}, 2003] for a review
and references therein). 
The most general form of the stress release model (SRM) reads
\be
\lambda_i(t) = \lambda_0
\exp \left[ b_i t  - b_i  \sum_j  g_{ij}  S_j(t)  \right]~.
\label{stressrelease}
\ee
In the SRM, the tectonic loading increases the stress at cell $i$ linearly 
in time according to $b_i t$ and earthquakes on this
cell and elsewhere relax (or load) the stress at cell $i$.
$S_j(t) = \sum_{\tau \leq t} \Delta \sigma_j(\tau)$ is the cumulative stress release
over all past earthquakes that occured in that cell $j$. 
$S_j(t)$ impacts region $i$ 
through the time-independent coupling (or stress transfer) coefficient $g_{ij}$.
In our model, the tectonic loading appears in contrast through 
the rate $\lambda_{\rm tec}({\vec r}, t)$. The SRM views the earthquakes
as mostly unloading this tectonic stress (of course stress load
is possible) while we view the earthquakes more symmetrically
as both promoting or shadowing the seismic activity elsewhere
around an average rate controlled by $\lambda_{\rm tec}({\vec r}, t)$. Finally, the
most important difference lies in the fact that the SRM
assumes an infinite time memory of past stress releases in the cumulative stress
release function $S_j(t)$, while we view the stress transfer by each past
earthquake as transient due to visco-elastic processes
in the crust and mantle. This infinite time memory of the SRM
is made necessary to compensate for the tectonic loading $b_i t$ in order
to obtain a statistically stationary process. In contrast, our model
is better devised to deal with transient stress perturbations induced by
past earthquakes. As a consequence, the SRM is not built to
produce aftershocks (see however the two-cells version of [{\it Borokov and Bebbington}, 2003],
which does produce first generation aftershocks obeying the Omori law).
The fits of seimic catalogs to the SRM indeed use declustered
data. Thus, the fundamental difference between our 
model (\ref{fund},\ref{funddis}) and the linked SRM (\ref{stressrelease})
is that the later does not describe either aftershocks or delayed
triggered seismicity (Imoto et al. [1999] introduced a fixed one-time delay in order
to produce periodic-type behavior).
Our model (\ref{fund},\ref{funddis}) is
an important generalization to account for the delayed triggering processes, 
which have been found to explain many phenomenological 
observations of seismicity [{\it Helmstetter and Sornette}, 2003a]. Actually,
all the results that we derive below derive from the time-dependent
memory kernel of our model, which are thus absent in the SRM.

\subsection{Reduction to time-only dynamics}

Starting from (\ref{fund}), we write the space-time kernel $g({\vec r}, t)$
in a separable form
\be
g({\vec r}, t) = f({\vec r}) \times h(t)~.
\label{mfgma}
\ee
This choice is made for the sake of simplicity and is in the spirit of 
the specification of the ETAS which also assumes separability of the bare
kernel in space and time. Helmstetter and Sornette [2002b]
have shown that the cascade of
triggering of events which are decoupled in time and space in their first generation 
eventually leads to a coupling in space and time corresponding
to a sub-diffusion process. Here, we expect a similar mechanism to operate
when taking into account all generations of earthquake triggering (a mainshock
generates aftershocks of first-generation, which themselves trigger 
aftershocks of second-generation and so on).

With the separable form (\ref{mfgma}), expression (\ref{fund}) can be transformed into
\be
\lambda({\vec r}, t) = \lambda_{\rm tec} ({\vec r}, t)  
~\exp \left[ \beta \int_{-\infty}^t d\tau
~s({\vec r}, \tau)  h(t-\tau) \right]~,
\label{fund2}
\ee
where
\be
s({\vec r}, \tau) = \int d{\vec r}~' ~\Delta \sigma({\vec r}~', \tau) ~f({\vec r} - {\vec r}~')
\label{sourmca}
\ee
is the effective source at time $\tau$ at point ${\vec r}$ resulting from all
events occurring in the spatial domain at the same time $\tau$. The
separable form of the kernel $g({\vec r}, t)$ in 
(\ref{mfgma}) allows us to study the time-dependent properties of the model,
independently of its space properties. Since expression (\ref{fund2}) is
defined for any ${\vec r}$, if we assume space homogeneity, or if we restrict
to a specific domain, we can drop the reference to ${\vec r}$ without loss
of generality and get
\be
\lambda(t) = \lambda_{\rm tec} (t)  ~
\exp \left[ \beta \int_{-\infty}^t d\tau
~s(\tau) ~ h(t-\tau) \right]~.
\label{fund3}
\ee
In contrast with the ETAS model in which the time-only equation of the
conditional Poisson rate describes the seismicity integrated over all
space, here the time-only equation (\ref{fund3}) refers to a specific location.
If space is homogeneous ($\lambda_{\rm tec} ({\vec r}, t)$ and $s({\vec r}, \tau)$
are independent of ${\vec r}$), then equation (\ref{fund3}) gives the 
conditional seismic Poisson intensity at any point.

\subsection{The distribution of stress source strengths}

The fact that the source is given by (\ref{sourmca}) 
should allow us to constrain its statistics.
Indeed, Kagan [1994] has suggested using theoretical calculations, simulations
and measurements of rotation of earthquake focal mechanisms that the stress
change in earthquake focal zones due to past earthquakes should follow the 
symmetric Cauchy distribution 
\be
L_1(x) = {1 \over \pi}~ {1 \over 1+x^2}~,
\label{mgbmls}
\ee
or perhaps even distributions decaying as power laws with even smaller exponents. The Cauchy
distribution is a stable L\'evy law with power law tail exponent $\mu=1$
and can be shown to be the distribution for the stress at any point due
to a random uniform distribution of sources mediated to that point
via the elastic Green function [{\it Zolotarev and Strunin}, 1971; {\it Zolotarev}, 1986]
(see also [{\it Sornette}, 2004], chap.~17, for a general presentation).
The physical mechanism for the Cauchy distribution is thus precisely the
summation (\ref{sourmca}) over earthquake sources $\Delta \sigma({\vec r}~')$ with stress transfer 
given by the elastic Green function $f({\vec r} - {\vec r}~')$ in the crust.
The Cauchy exponent $\mu=1$ is obtained for a uniform spatial
distribution of sources in 3D with the elastic Green function $\sim 1/r^3$ in 3D,
or for a uniform spatial
distribution of sources in 2D with the elastic Green function $\sim 1/r^2$ in 2D. For
sources on a fractal with dimension $D_f \leq 3$, $\mu = D_f/3$ in 3D [{\it Kagan}, 1994].

However, it must be kept in mind that the large values of the stress sources $s$
that contribute to the slow power law decay of the Cauchy distribution
result from the assumption that earthquakes are point-wise 
such that a probe put at random in the medium can come arbitrarily close
to these sources: it is the divergence
$\sim 1/r^3$ (in 3D) of the stress field close to such a singular source which 
is at the origin of the Cauchy distribution (see Chap.~17 of [{\it Sornette}, 2004]). 
In reality, such singular
power law behavior transforms into a much weaker $1/\sqrt{r}$ singularity close to crack tips
and then crosses over to a smooth behavior due to plasticity and damage that
smooth out the singularity sufficiently close to the fault edges. 

To capture in a phenomenological way these features, we will use a 
power law distribution of the source strengths
\be
P(s) \approx {C \over (s^2+s_0^2)^{1+\mu \over 2}} \sim {C \over s^{1+\mu}}
\label{vcmnfdlqa}
\ee
where the scale factor $C$ is given by $C \propto s_0^{\mu}$ and
$s_0$ is a characteristic scale proportional to the average stress drop.
The value $\mu=1$ recovers the special case
of the Cauchy distribution advocated by Kagan [1994].
Notice that the distribution of the source strengths is symmetric, implying
that the effect of an earthquake in the past can either enhance or shadow
the present seismicity. 
This generalizes the ETAS model as both 
stress triggering and stress shadowing are taken into account symmetrically
while the ETAS model describes only stress triggering.

\subsection{Time-dependent stress relaxation kernel}

The last ingredient we need to specify is the time dependence of the 
memory kernel $h(t)$. We postulate a stress relaxation function
\be
h(t) = {h_0 \over (t+c)^{1+\theta}}~, ~~~~{\rm for}~ 0< t \leq T~,
\label{f,ma,fa}
\ee
which is of the Omori form with the usual small time-scale cut-off $c$. 
To ensure convergence of the correlation function of
deterministic processes with memory governed by $h(t)$ for any possible
values of $\theta$, we truncate the power law in
(\ref{f,ma,fa}) at some large time $T$, which we
call the ``integral time scale:''  it is the largest time scale up to which 
the memory of a past event survives. $T$ can thus be interpreted as the effective
Maxwell time of the relaxation process. The sharp cut-off implied by $T$ is
invoked for convenience and can be replaced by a smooth cross-over
using for instance an exponential roll-off of the form
\be
h(t) = {h_0 \over (t+c)^{1+\theta}}~e^{-t/T}~,
\label{f,ma,faTT}
\ee
without changing our main conclusions below.

Just after an event over a time scale slightly larger than the 
time for the propagation and attenuation of dynamical stress waves,
that we note $t=0^+$ for short,
the stress equilibrates to its static value and we should have
\be
h(0^+) = {h_0 \over c^{1+\theta}} = 1~,
\label{mngmle}
\ee
to express that the static stress has not had time to relax yet. 

Expressions (\ref{f,ma,fa}) or (\ref{f,ma,faTT}) 
can be rationalized from the time dependence of the visco-elastic
Green function in 1D which gives $\theta =-1/2$, 
in 2D which gives $\theta=0$ or in 3D which gives $\theta=1/2$.
In fact, a better formulation will require a space-time dependence
of the evolution of the stress field.
Another argument is to view to local stress as proportional
to the local strain rate, which is itself proportional to the local microscopic
seismic rate which obeys the Omori law. From a micro-mechanical point of view,
such slow relaxation process (\ref{f,ma,fa}) are associated with 
dislocation motion, stress corrosion and hydrolytic weakening processes
[{\it Sornette}, 1999]. We also would like to underline that 
expression (\ref{f,ma,fa}) implies the absence of a well-defined
characteristic time scale for $c<t<T$ where $c \ll T$, and embodies the 
complex non-Maxwell relaxation processes in the crust, its coupling with the lower
visco-elasto-plastic crust and more ductile upper mantle. The coexistence of 
many different time scales can be captured by such power law decay.

Actually, there is a more profound origin of the dependence (\ref{f,ma,fa})
of the stress relaxation. Assume that an earthquake loads a given region
according to the elastic stress redistribution that can be estimated using 
standard methods of elasticity [{\it King et al.}, 1994; {\it Stein}, 2003].
Then, by the activation processes discussed in section \ref{ss1}, this stress
redistribution induced by the ``mainshock'' will give rise to an increase
of seismicity at that region. The triggered earthquakes will lead
to new sources of stress redistribution, which themselves modify the stress field,
tending to decrease it. This process gives rises to a slow power law
relaxation of the stress field [{\it Lee and Sornette}, 2000]. The power law
decay embodied in (\ref{f,ma,fa}) can thus be viewed as resulting from the 
microscopic process of stress redistribution and relaxation which occur below
the scale of observation. There is no reason for the physics to change 
and this law (\ref{f,ma,fa}) is an effective renormalization of many microscopic
relaxation processes.

\subsection{Summary of the model definition}

Summarizing, in a discrete form, our model reads
\be
\lambda(t) = \lambda_{\rm tec} ~
\exp \left[ \beta \sum_{i~|~t_i\leq t} 
~s(t_i) ~ h(t-t_i) \right]~,
\label{fundfsaf}
\ee
where the stress at time $t$
is the sum of the contributions over all previous earthquakes 
that occurred at times $t_i \leq t$, with stress sources given by the 
power law distribution (\ref{vcmnfdlqa}) and with a power law time-dependent
stress relaxation kernel (\ref{f,ma,fa}).
In the sequel, we take a constant seismic rate $\lambda_{\rm tec}$ in the 
absence of stress perturbation $s(t_i)=0$. 

It is convenient
to rewrite (\ref{fundfsaf}) as
\be
\lambda(t) = \lambda_{\rm tec} ~ e^{\beta \omega(t)}~,
\label{vmjhi}
\ee
where 
\be
\omega(t) = \sum_{i~|~t_i\leq t}  ~s(t_i) ~ h(t-t_i)~.
\label{mgkmkhy}
\ee

Model (\ref{fundfsaf}) belongs to the class of nonlinear self-excited point-processes,
in which the nonlinear function is the exponential in our case 
[{\it Br\'emaud and Massouli\'e}, 1996; {\it Br\'emaud et al.}, 2002].
Br\'emaud et al. [2002] have given the general condition guaranteeing the existence 
of a statistically stationary solution in the case of unbounded nonlinear 
function, sub-exponential distribution of the marks (stress changes $s$'s) and long
memory kernel as is our case. In our model, the introduction of the integral time scale $T$
in (\ref{f,ma,fa}) ensures the existence of the correlation function for
any process and for any values of the exponent $\theta$.

\section{Derivation of the magnitude dependence of Omori's law}

In this section, our goal is to derive Omori's law from model (\ref{fundfsaf})
with (\ref{vcmnfdlqa}) and (\ref{f,ma,fa}).
The problem can be formulated as follows. Omori's law quantifying the
decay of seismic activity after a ``mainshock'' occurring
at the origin of time amounts to determining
the typical time dependence of $\lambda(t)$ conditioned on a value $\lambda_M$
realized at $t=0$ which is larger than average. This is due to the fact
that a mainshock of magnitude $M$ induces a local burst of seismic activity
proportional to $K~10^{\alpha M}$, where $K$ and $\alpha$ are two positive constants
[{\it Helmstetter}, 2003]. We note however that previous determinations
of the productivity exponent $\alpha$ have assumed the constancy of $p$
with mainshock magnitude; in the presence of an exponent $p(M)$ which 
increases with $M$ as we find here, past values of $\alpha$ have probably
be underestimated.

\subsection{Theory of Omori's law by generalization 
of conditional expectations of seismic rates to power law}

In the case where averages exist, the Omori law can be expressed
in the following general form:
\be
{\rm E}[\lambda(t)|\lambda_M] = \lambda_{\rm tec} {\rm E}[e^{\beta \omega(t)}|\omega_M]~,
\label{mgmdlsas}
\ee
where ${\rm E}[.]$ denotes the mathematical expectation or average of ensemble
of equivalent statistical realizations of the process.

In the case where averages and variances and covariances
are ill-defined mathematically, as is the case 
for Cauchy distributions and for power laws with $\mu \leq 2$, 
a typical measure of conditional seismicity rate can
be defined at any quantile level $q$ by the probability ${\rm Pr}[\lambda(t)>\lambda_q|\lambda_M]$
that the rate $\lambda(t)$ be larger than the quantile $\lambda_q$ conditioned on the
fact that the seismic rate was at some given value $\lambda_M$ at time $0$:
\be
{\rm Pr}[\lambda(t)>\lambda_q|\lambda_M] = 
{\rm Pr}[e^{\beta \omega(t)}> {\lambda_q \over \lambda_{\rm tec}}|\omega_M]
={\rm Pr}[\omega(t)> (1/\beta)\ln \left({\lambda_q \over \lambda_{\rm tec}}\right)|\omega_M]~.
\label{mgmdlsasdss}
\ee
We are interested in monitoring the time evolution
$\lambda_q(t)$ of the seismic rate quantile at some probability level $q$ (which can 
be varied to explore different fluctuation levels).

If the source terms $s(t_i)$ were centered Gaussian random variables, $\omega$ would also
be normally distributed. Using (\ref{mgmdlsas}), this would allow us to obtain
\be
{\rm E}[e^{\beta \omega(t)}|\omega_M] = \exp \left[ \beta {\rm E}[\omega(t)|\omega_M] 
+ {\beta^2 \over 2} {\rm Var}[\omega(t)|\omega_M] \right]~,
\ee
where
\be
{\rm E}[\omega(t)|\omega_M] = \omega_M~{{\rm Cov}[\omega(t),\omega_M] \over {\rm Var}[\omega_M]}~.
\label{mgmel}
\ee
Using the definition (\ref{mgkmkhy}), this would provide a closed formed expression for 
the Omori law describing the relaxation of the conditional rate ${\rm E}[\lambda(t)|\lambda_M]$.
The physical meaning of (\ref{mgmel}) is that one can write a linear regression
\be
\omega(t) = \gamma(t) \omega_M + \epsilon~,
\label{mgnslwf}
\ee
where $\gamma(t)$ is a non-random factor and $\epsilon$ is a centered Gaussian noise
with zero correlation with $\omega_M$. Equation (\ref{mgnslwf}) writes that
the best predictor of $\omega$ given $\omega_M$ is $\gamma \omega_M$,
i.e., ${\rm E}[\omega(t)|\omega_M] = \gamma \omega_M$ with 
\be
\gamma =
{{\rm Cov}[\omega(t),\omega_M] \over {\rm Var}[\omega_M]}~,
\label{mfngzla}
\ee
which retrieves (\ref{mgmel}).

However, as we explained above, the sources are distributed according to a
distribution (\ref{vcmnfdlqa}) which can be expected to have
a heavy tail, such that both
its variance and average are not mathematically defined, or if they are defined
converge very poorly to their asymptotic values even in large data sets. Thus, the standard 
statistical tools
of expectation, variance and covariance can not be used and we need a completely novel
approach to tame mathematically these wild fluctuations. For this, we use the 
insight that the natural generalization of the variance for power laws
$p(x) \approx C/x^{1+\mu}$ with infinite
variance (i.e., with $\mu<2$) is the scale parameter $C$, as it possesses
most of the properties of the variance for Gaussian random variables: it is 
additive under convolution of the distribution and it replaces the variance in the
expression of the characteristic function of the distribution (see Chap.~4 of 
[{\it Sornette}, 2004]). 

In the power law case, due to the linear form of (\ref{mgkmkhy}),
we can still write (\ref{mgnslwf}) 
but with $\omega(t), \omega_M$ and $\epsilon$
being power law distributed random variables with the same exponent $\mu$
and with scale factors equal respectively to $C_{\omega}$ (for
$\omega$ and $\omega_M$) and $C_{\epsilon}$.
The key idea is that $\gamma$ can be determined by a generalization of 
(\ref{mfngzla}), involving generalizations of the covariance and variance. 
This generalization consists in forming the random variable defined as the 
product $\omega \omega_M = \gamma \omega_M^2 + \epsilon \omega_M$. It
is straightforward to show that the distribution of $\omega \omega_M$ consists of two
main contributions, (i) a dominant power law with exponent $\mu/2$ and scale
factor $C_{\omega \omega_M} = \gamma^{\mu/2}~C_{\omega}$, and (ii) a sub-dominant power law with 
exponent $\mu$ (with a logarithmic correction) and scale factor $C_{\omega}C_{\epsilon}$.
This has the following practical implication: if one measures or calculates
the leading power law decay of $\omega \times \omega_M$, the measure of its
scale factor gives access to the parameter $\gamma$ through the expression
\be
\gamma(t) = \left( C_{\omega \omega_M} \over C_{\omega} \right)^{2 \over \mu}~,
\label{mgmel2}
\ee
where the time dependence of $\gamma(t)$ comes from that of $C_{\omega \omega_M}$
as we show below.

This expression (\ref{mgmel2}) generalizes the standard result (\ref{mfngzla}):
notice that the case $\mu=2$ recovers (\ref{mfngzla}) 
with the correspondence $C_{\omega}= {\rm Var}[\omega]$ and 
$C_{\omega \omega_M}={\rm Cov}[\omega(t),\omega_M]$. This is expected since, as we
said above, the scale factor reduces to the variance for the Gaussian distribution
and the stable L\'evy law with exponent $\mu=2$ turns out to be nothing but the Gaussian law!
We note that this method consisting of generalizing the covariance by introducing the concept
of ``tail-covariance'' has been previously used to extend the Kalman filter of
data assimilation to power law distributed noise processes [{\it Sornette and Ide}, 2001].

Using (\ref{mgkmkhy}), we form the product 
\be
\omega(t) \omega_M =  \sum_{i~|~t_i\leq t} \sum_{j~|~t_j\leq 0}  
~s(t_i)~s(t_j) ~ h(t-t_i)~h(-t_j)~, 
\label{mgkmaaakhy}
\ee
where the $s$'s are random variables with power law tail with exponent $\mu$
(specifically, in this paper, the $s$'s are Cauchy variables with $\mu=1$ but
we give the derivation for any value of $\mu$).
Then, using standard calculations (see Chap.~4 of [{\it Sornette}, 2004]), 
the terms in the double sum in (\ref{mgkmaaakhy}) that contribute to the 
leading asymptotic power law tail with exponent $\mu/2$ correspond to the 
diagonal terms $i=j$, while all the other terms contribute to the sub-leading
power law tail with exponent $\mu$ with logarithmic corrections. This gives
the expression of the scale factor $C_{\omega \omega_M}^{\{\mu/2\}}$
of the dominating power law with exponent 
$\mu/2$
\be
C_{\omega \omega_M}^{\{\mu/2\}} = C_{\omega} \sum_{i~|~t_i\leq 0} 
\left[h(t-t_i) h(-t_i)\right]^{\mu \over 2}~.
\label{mgmg,sls}
\ee
Together with (\ref{mgmel2}), this yields
\be
\gamma = \left( \sum_{i~|~t_i\leq 0} 
\left[h(t-t_i) h(-t_i)\right]^{\mu \over 2} \right)^{2 \over \mu}~.
\label{nbnsjs}
\ee
Since $h$ is dimensionless, $\gamma$ is also dimensionless, as it should from
its definition (\ref{mgnslwf}).

In order to perform a theoretical analysis of (\ref{nbnsjs}), it is 
convenient to transform the discrete sum into a continuous one. 
Let us consider the times $t_i$'s in the sum $\sum_{i~|~t_i\leq 0}$ in 
(\ref{nbnsjs}). Ideally, these times $t_i$ should themselves be 
determined self-consistently and are known to follow on average
an inverse Omori law. However, such an inverse Omori law is a statistical
property observed only for a large ensemble of stacked foreshock sequences
while individual sequences exhibit approximately constant seismic rates
[{\it Helmstetter and Sornette}, 2003a]. In order to simplify the analysis,
we thus approximate the seismicity prior to a mainshock as being
approximately constant in time and uniform in space. This allows us to
introduce the average time interval $\Delta t$ between two events preceding
the mainshock. We expect $\Delta t$ to be of the order of the 
small time-scale cut-off $c$ in (\ref{f,ma,fa}). This is natural
if the stress relaxation is in significant part due to seismicity itself:
then, $c$ is the time scale beyond which the stress relaxation 
starts to be felt, i.e., when new earthquakes occur in the vicinity of the 
source. 
Then, we can approximate the discrete sum as follows:
\be
\sum_{i~|~t_i\leq 0}  \to  \int_{-\infty}^0  {dt \over \Delta t}~.
\ee
Using the form (\ref{f,ma,fa}) for the stress relaxation kernel gives
in continuous time after some manipulations
\be
\gamma(t) = {h_0^2 \over \Delta t^{2/\mu}}
\left( {1 \over t^{(1+\theta)\mu -1}}\int_0^{(T/t)-1} dx {1 \over (x+1+(c/t))^{(1+\theta)\mu/2}}~
{1 \over (x+(c/t))^{(1+\theta)\mu/2}} \right)^{2 \over \mu}~,
\label{mgmlss}
\ee
where $h_0 = c^{1+\theta}$ according to (\ref{mngmle}).
We verify that $\gamma(t)$ is dimensionless as it should.
Under the change of variable $x \to y=x+{c \over t}$, expression (\ref{mgmlss}) can be written
\be
\gamma(t) = {h_0^2 \over \Delta t^{2/\mu}}  \left( {1 \over t^{2m-1}}
\int_{c/t}^{{T+c \over t}-1} dy {1 \over (y+1)^{m}}~
{1 \over y^{m}} \right)^{2 \over \mu}~,
\label{mgmlssaafa}
\ee
where $m = (1+\theta)\mu/2$.

We now have all the ingredients to estimate (\ref{mgmdlsasdss}).
We thus obtain
\be
{\rm Pr}[\omega(t)>y|\omega_M] = {\rm Pr}[\gamma \omega_M +\epsilon>y|\omega_M]
= {\rm Pr}[\epsilon>y-\gamma \omega_M|\omega_M] = {\bar F}(y-\gamma(t) \omega_M)~,
\label{mgmgkes}
\ee
where ${\bar F}(\epsilon)$ is the complementary cumulative distribution of $\epsilon$.
Using (\ref{mgmgkes}) in (\ref{mgmdlsasdss}), this leads to
\be
{\rm Pr}[\lambda(t)>\lambda_q|\lambda_M] = 
{\bar F}\left((1/\beta)\ln \left({\lambda_q \over \lambda_{\rm tec}}\right)-\gamma(t) \omega_M\right)~.
\label{mgmdlsasaadss}
\ee
The typical time evolution of the seismicity rate $\lambda(t)$ conditioned
on the rate $\lambda_M$ at time $0$ is thus given by fixing the quantile probability
to some level ${\rm Pr}[\lambda(t)>\lambda_q|\lambda_M]=q$, leading to
\be
{1 \over \beta} \ln \left({\lambda_q \over \lambda_{\rm tec}}\right)-\gamma(t) 
\omega_M = {\bar F}^{-1}(q)~,
\ee
or equivalently
\be
\lambda_q(t) = A_q ~\lambda_{\rm tec}~e^{\beta \gamma(t) \omega_M}~,
\label{gnnlblwd}
\ee
where $A_q = \exp \left( \beta {\bar F}^{-1}(q) \right)$.
The time-dependence of the seismic decay rate is thus determined by 
(\ref{gnnlblwd}), which requires the determination of the time-dependence of 
$\gamma(t)$ given by (\ref{mgmlss}) (and more generally by (\ref{nbnsjs})).

In the case where the stress relaxation is a pure exponential $h(t) \sim e^{-t/T}$,
then using (\ref{mgmlss}), we find $\gamma(t) \sim e^{-t/T}$ also and the seismic rate 
$\lambda_q(t)$ given by (\ref{gnnlblwd}) also relaxes exponentially to a constant
background. The novel effect that we describe below comes from the interplay between
the exponential thermal activation process and the long-memory process of the stress relaxation.

Note that this approach holds also for $\mu>2$ for which variances and 
covariances exist, which can allow the application of (\ref{mfngzla}).
However, using this power approach provides more robust estimators of the
typical values of seismic rates when the convergence of the mean and of variances
are very slow, as occurs for power laws (see Chap.~3 of [{\it Sornette}, 2004]).

\subsection{Study of the predicted seismicity rate for different values
of the two key exponents $\mu$ and $\theta$}

Our purpose is to show that, for a rather broad range of values of the exponents
$\mu$ and $\theta$ defining the model, $\lambda_q(t)$ is approximately given by
\be
\lambda_q(t) \sim {1 \over t^{p(M)}}~,
\label{mgkhh}
\ee
with 
\be
p(M)= a \beta M + b \beta~,
\label{mgffz}
\ee
where $a >0$ and $M$ is the mainshock magnitude. In a nutshell, expression (\ref{mgkhh}) with
(\ref{mgffz}) result from the interplay between the heavy-tailed distribution of 
stress perturbations and the long time memory of the stress relaxation on the one hand
and the exponential 
dependence of the seismic rate on the stress field on the other hand.

\subsubsection{Case $\theta=-1/2$ and $\mu=2$ (or with a Gaussian distribution of stress source
strengths) \label{mrwperf}}

This interplay between the long time memory of the stress relaxation and the
exponential dependence of the seismic rate on the stress field is exemplified 
by the case $\theta=-1/2$ and $\mu=2$, which has previously been shown to give
exact multifractal properties in the time domain [{\it Muzy and Bacry}, 2002], and as a
consequence continuous dependence of the relaxation exponents $p(M)$ as a function
of shock magnitudes $M$.
This continuous dependence of the exponent $p(M)$ has actually been documented
empirically in this case in another context of aftershock decay following shocks
in financial markets [{\it Sornette et al.}, 2003]. 
Schmitt [2003] has studied in details the properties of the process (\ref{vmjhi}) with 
(\ref{mgkmkhy}), when $s(t)$ is a Gaussian white noise (corresponding formally to $\mu=2$),
with the memory exponent $\theta=-1/2$.

An intuitive grasp of this behavior is obtained by examining $d\gamma/dt$:
\be
{d \gamma \over dt} = - {(h_0^2 /\Delta t^{2/\mu}) \over t}~\left[
{(T+c)^{1/2} \over (T+c-t)^{1/2}} - {c^{1/2} \over (t+c)^{1/2}}\right]~.
\label{gmmgmrl}
\ee
It is easy to see that the first (resp. second) term of the bracket in the r.h.s. is always larger
(resp. smaller) than $1$, which ensures that ${d \gamma \over dt}$ is always negative.
In addition, for $t<T$, the bracket in the r.h.s. is almost constant and close to $1$, showing
that $d\gamma/dt \approx -1/t$ and thus $\gamma(t) \approx {\rm constant}_1 - 
{\rm constant}_2  \times \ln (t/T)$. Then expression (\ref{gnnlblwd})
leads to (\ref{mgkhh},\ref{mgffz}) using the fact that 
$\omega_M \propto \ln (\lambda_M) \propto \ln (K~10^{\alpha M}) =\alpha \ln 10 ~M + \ln K$,
i.e., $\omega_M$ is linearly related to the magnitude $M$. Here, we have used
the productivity law that an earthquake of magnitude $M$ produces on average
a number of aftershocks proportional to the exponential of its magnitude 
($\alpha$ is the productivity exponent often reported between $0.5$ and $1$).

\subsubsection{Condition $2m=\mu (1+\theta) = 1$}

\begin{figure}
\begin{center}
\psfig{file=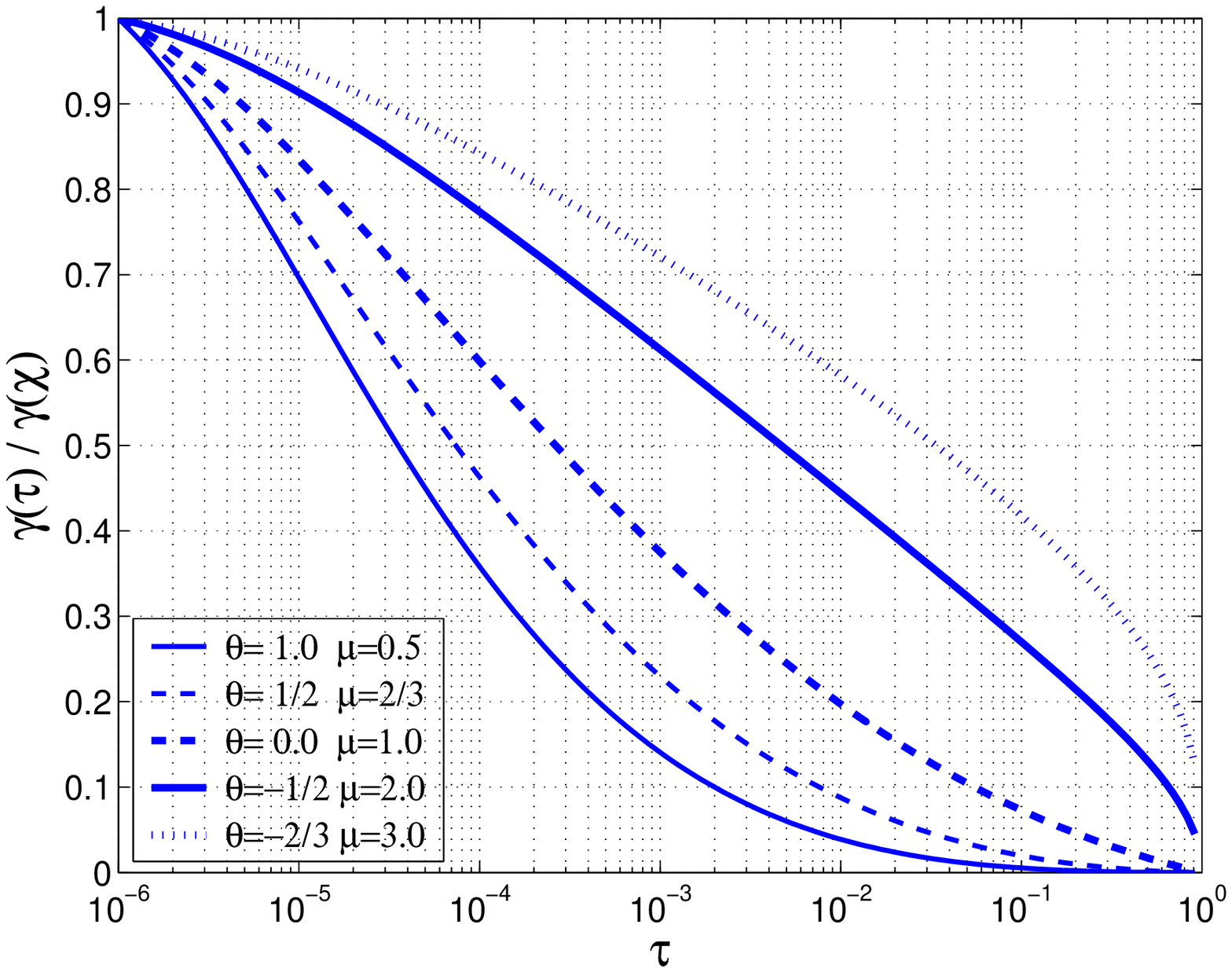,width=14cm}
\caption{\label{figmutheta1} Numerical
evaluation of the normalized
$\gamma(\tau)/\gamma(\chi=c/T)$ (linear scale) defined by (\ref{mgmlssaafa})
as a function of $\tau =t/T$ (logarithmic scale), for several pairs $(\mu, \theta)$ which
obey the condition $\mu (1+\theta) = 1$ exactly, using $c/T=10^{-6}$.
We verify the existence of a linear decay of $\gamma(t)$ in the variable
$\ln t$, which qualifies an Omori power law (\ref{mgkhh}). Since the logarithm of
the seismicity rate $\lambda(t)$ is proportional to $\gamma$ (with 
a coefficient of proportionality involving the inverse temperature and 
the magnitude of the mainshock), a linear behavior
qualifies a power law decay of the seismic rate, whose range is rather sensitive
to the temperature and the ratio $c/T$.
}
\end{center}
\end{figure}

This case corresponds to taking the exponent $m = (1+\theta)\mu/2$ defined in (\ref{mgmlssaafa})
equal to $1/2$. Then, ${d \gamma^{\mu/2} \over dt}$ has exactly the expression
given by the right-hand-side of (\ref{gmmgmrl}), showing that ${d \gamma^{\mu/2} \over dt}$
is close to $-1/t$, and thus $\gamma^{\mu/2}(t)  \approx {\rm constant}_1 - 
{\rm constant}_2 \ln (t/T)$ which, for not too small nor too large $t$'s and 
for ${\rm constant}_1 < {\rm constant}_2$, gives
$\gamma^(t) \approx {\rm constant}'_1 - 
{\rm constant}'_2  \times \ln (t/T)$. This yields (\ref{mgkhh},\ref{mgffz}).
Typically, the power law behavior is observed over more than two decades in time,
which is comparable to empirical observations. Figure \ref{figmutheta1} shows the numerical
evaluation of $\gamma(t)$ as a function of $\ln (t/T)$ for several pairs $(\mu, \theta)$ which
obey the condition $\mu (1+\theta) = 1$ exactly:
$(\mu=3, \theta=-2/3); (\mu=2, \theta=-0.5); (\mu=1, \theta=0);
(\mu=2/3, \theta=0.5); (\mu=0.5, \theta=1)$, using $c/T=10^{-6}$.
We verify the existence of a linear decay of $\gamma(t)$ in the variable
$\ln t$, which qualifies an Omori power law (\ref{mgkhh}).
The range of this Omori power law regime depends rather sensitively on the temperature,
on the mainshock magnitude and on the ratio $c/T$ since the logarithm of the seismic rate is
proportional to $\gamma$ via a coefficient of proportionality involving the
inverse temperature and the mainshock magnitude, as shown in (\ref{gnnlblwd}).
Then, by the same argument as in section \ref{mrwperf}, the linear dependence of
$\omega_M$ on $M$ in expression (\ref{gnnlblwd})
leads to (\ref{mgkhh},\ref{mgffz}).

The fact that $\gamma(t)$ is asymptotically exactly logarithmic in time
for $2m=\mu (1+\theta) = 1$ and thus that the seismic rate $\lambda(t)$ is
an Omori power law can be recovered from a different construction motivated 
by multiplicative cascades introduced in turbulence. The 
discrete multiplicative cascade model of hydrodynamic turbulence can be
extended in distribution to the continuous limit [{\it Schmitt and Marsan}, 2001] 
and shown to take the form (\ref{mgkmkhy}) expressed with the discrete sum
replaced by a continuous integral. Then, Schmitt and Marsan [2001] show that 
the only condition for $\exp[\omega]$ (in our case $\lambda$ through (\ref{vmjhi}))
to be ``logstable multifractal'' is that the exponent $\mu$ of the
distribution of the innovations $s$ and
the exponent $1+\theta$ of the memory kernel be related by the condition
$\mu (1+\theta) = 1$ that we have derived above through a different route.
In a nutshell, their argument (which applies in distribution to one-point
statistics but not to multi-point statistics nor in process) is as follows.
The logarithm of $\lambda$ is constructed as an integral in the time-scale $(t,a)$ domain
within a cone with apex on the time axis
of identically independent distributed innovations distributed according to 
a L\'evy distribution with exponent $\mu$. The natural scale-invariant measure
for this integral
in this time-scale domain is $da dt/a^2$ in
the sense that it is (left-)invariant by the translation-dilation group
[{\it Muzy and Bacri}, 2002]. The contribution
of a vertical strip of width $dt = \ell$ centered at time $t-\tau$ (i.e.,
at a distance $\tau$ from the present time $t$) within the cone
is $\epsilon_1 [\ell \int_{t-\tau}^{+\infty} da/a^2]^{1/\mu} = \epsilon_1 (\ell/(t-\tau))^{1/\mu}$, 
where $\epsilon_1$
is a L\'evy distributed noise with scale factor unity. The exponent $1/\mu$ comes from the
dependence of the scale factor of sum of the $\ell/(t-\tau)$ L\'evy variables that
contribute to each trip. Now, in distribution, 
the total integral is thus equal to the sum of all strips up to time $t$, which gives
$\int^t d\tau~ \epsilon_1(\tau) (\ell/(t-\tau))^{1/\mu}$. This expression
is exactly of the form (\ref{mgkmkhy}) with $1+\theta=1/\mu$. QED.

\subsubsection{Case $2m=\mu (1+\theta) \neq 1$}

It is useful to study the generalization of (\ref{gmmgmrl}), which reads
\be
{d [t^{2m-1}\gamma^{\mu/2}] \over dt} = - {(h_0^2 / \Delta t^{2/\mu}) \over t^{2-m}}~\left[
{(T+c)^{1-(m/2)} \over (T+c-t)^{m/2}} - {c^{1-(m/2)} \over (t+c)^{m/2}}\right]~.
\label{gmmgmrafal}
\ee

\begin{figure}
\begin{center}
\psfig{file=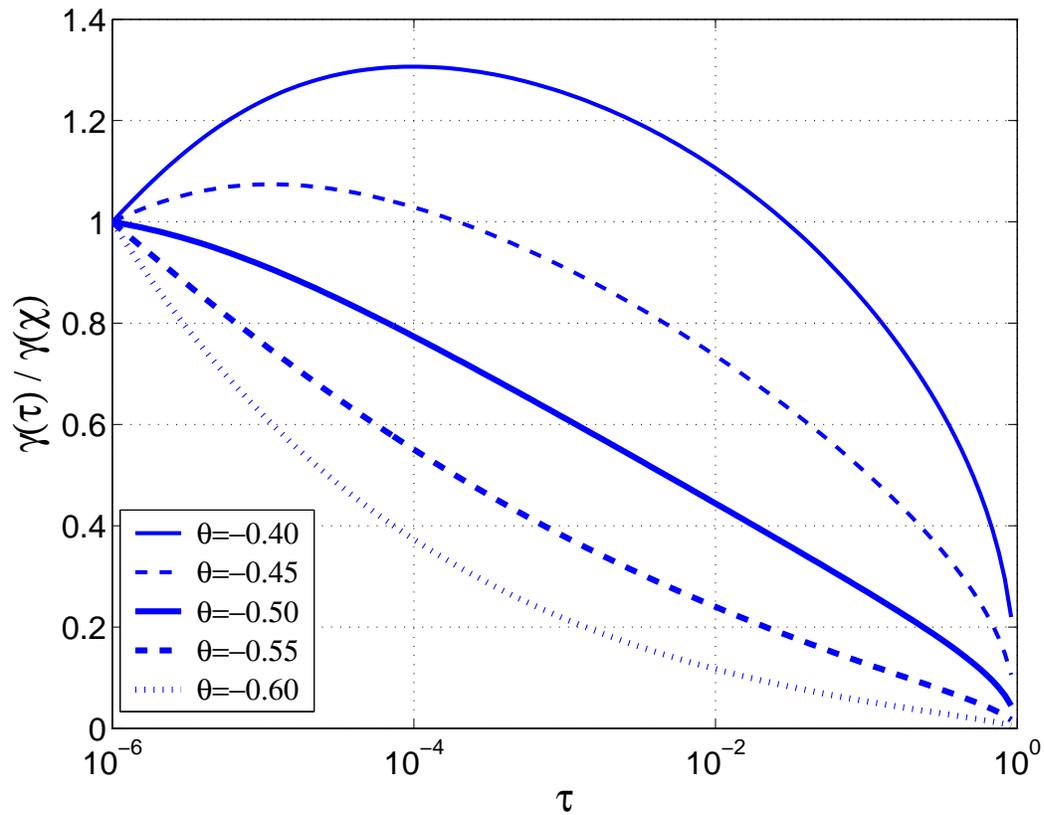,width=14cm}
\caption{\label{figmutheta2} Same as Fig.~\ref{figmutheta1}
for several values of $\theta$ at fixed $\mu=2$, to show the sensitivity of 
the existence of a linear behavior of $\gamma(t)$ as a function of $\ln t$
for values of $(\mu, \theta)$ which depart from the condition $\mu (1+\theta) = 1$.
Depending on the temperature and mainshock magnitude, a power law regime
for the seismic rate can be observed approximately over several decades even 
for values of  $(\mu, \theta)$ which depart from the condition $\mu (1+\theta) = 1$.
}
\end{center}
\end{figure}

Two cases must be distinguished. 
\begin{itemize}
\item For $2m=\mu (1+\theta)<1$, 
the first (resp. second) term of the bracket in the r.h.s. of (\ref{gmmgmrafal}) is always larger
(resp. smaller) than $1$, which ensures that 
${d [t^{2m-1}\gamma^{\mu/2}] \over dt}$ is always negative.
In addition, for $t<T$, the bracket in the r.h.s. is almost 
constant and close to $(T+c)^{1-m}$, showing
that $d[t^{2m-1}\gamma^{\mu/2}]/dt \sim -1/t^{2-m}$ and thus 
$\gamma^{\mu/2}(t) \approx {\rm constant}_1 \times (T/t)^{2m-1} + 
{\rm constant}_2 \times (T/t)^{m}$. $\gamma$ is thus a convex decreasing function of 
$\ln t$ (with a upward curvature) as shown in figure \ref{figmutheta2}. The mathematical form of the 
decay rate tends to slow down compared with a standard Omori power law. This convexity
tends to linearity as $m \to 1/2$.

\item  For $2m=\mu (1+\theta)>1$, we have $1-(m/2) < m/2$. As a consequence, for small $t$'s,
$(T+c-t)^{m/2} > (T+c)^{1-(m/2)}$ while the reverse inequality holds true for large $t$'s.
This reasoning shows that the bracket is negative for small $t$'s and becomes positive
for large $t$'s. Therefore, ${d [t^{2m-1}\gamma^{\mu/2}] \over dt}$ is positive for small $t$'s,
vanishes at an intermediate time and becomes negative for large $t$'s. This translates
into $\gamma$ increasing for small $t$'s,
passing through a maximum before decreasing for large $t$'s, as shown in figure \ref{figmutheta2}.
In this case, we can often observe an approximate linear decay of $\gamma(t)$ as
a function of $\ln t$, over two to three order of magnitudes in time
in the decaying part beyond the maximum, all the more so, the closer $m$ is to one.
Over this range, by the same argument as in section \ref{mrwperf}, the linear dependence of
$\omega_M$ on $M$ in expression (\ref{gnnlblwd})
leads to (\ref{mgkhh},\ref{mgffz}).
\end{itemize}

\section{Empirical investigation of Omori's law conditioned on mainshock magnitudes}

\subsection{Data selection and analysis of completeness}

We use the Southern Californian earthquakes
catalog with revised magnitudes (available from the Southern California 
Earthquake Center) as it is among the best available in terms of quality
and time span. Magnitudes $M_L$
are given with a resolution of $0.1$ from 1932 to 2003, in a region
from approximately $32^\circ$ to $37^\circ$N in latitude, and from $-114^\circ$
to $-122^\circ$ in longitude.

\begin{figure}
\begin{center}
\psfig{file=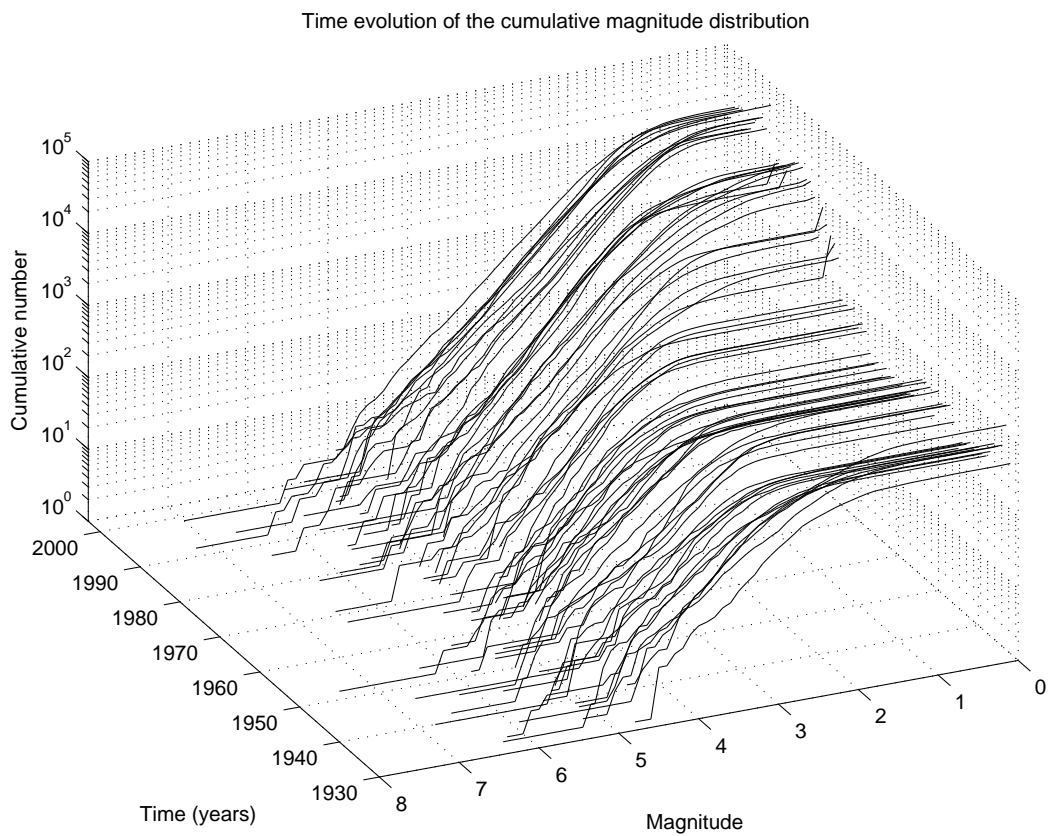,width=14cm}
\caption{\label{F1} Time evolution of the yearly cumulative 
magnitude distribution from $1932$ to $2003$ included, 
obtained from the Southern Californian earthquakes
catalogue with revised magnitudes (available at the Southern California 
Earthquake Center). Magnitudes $M_L$
are given with a $0.1$ resolution from 1932 to present, in a zone
roughly comprised within $32^\circ$ to $37^\circ$N in latitude, and within $-114^\circ$
to $-122^\circ$ in longitude.
}
\end{center}
\end{figure}

\begin{figure}
\begin{center}
\psfig{file=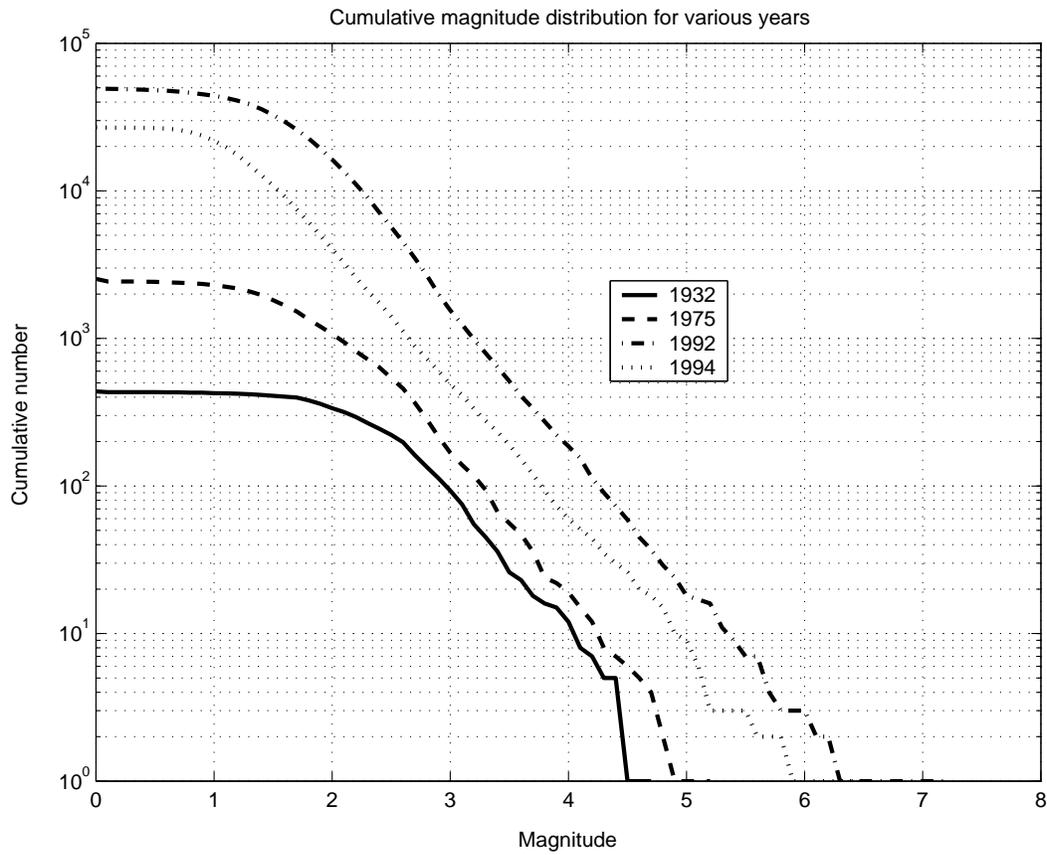,width=14cm}
\caption{\label{F2} Cumulative
magnitude distribution (CMD) for different time spans of the SCEC catalogue
used to define the approximate lowest magnitude $M_L$ of completeness. The CMD is 
approximately linear for $M_L>3$ for the whole lifespan
of the catalog, while it is linear
for all magnitudes larger than $2.5$ (respectively $2$ and $1.5$) for shocks
after $1975$ (respectively after $1992$ and $1994$). 
}
\end{center}
\end{figure}

Using all the data of this catalog, we compute the complementary cumulative 
magnitude distribution for each year from $1932$ to $2003$ included. This
gives 72 Gutenberg-Richter distributions which are shown in
Figure \ref{F1}. The logarithms of the distributions as a function of 
magnitude are approximately linear for the largest magnitudes and exhibit
a cross-over to a plateau at small magnitudes. In order for our analysis
to be robust, we need to address the question of completeness of the catalog.
The level of low-magnitude plateau 
increases with time, which is the consequence of the evolution of the
seismic network: as more stations are added, the spatial coverage increases,
so that the number of events with fixed magnitude also increases. One can
also observe that events of smaller and smaller magnitudes are 
detected and located when going from 1932 to the present. This also 
leads to an increase of the number of events recorded in the catalog.
As the typical time scale over which we analyze the Omori law decay 
is about $1$ year, which is small compared with the average time scale of
the evolution of the network coverage (see Figure \ref{F1}), 
we will consider all events in the catalog that have a magnitude that belong to the
linear part of the magnitude distribution curve. Figure \ref{F2}) shows
several complementary cumulative distributions of earthquake magnitudes
for the four years 1932, 1975, 1992 and 1994. Taking the linear relationship
between the logarithm of the distribution as a function of magnitude 
as the standard criterion for completeness [{\it Kagan}, 2003], we infer that the catalog is
approximately complete for $M_L>3$ in 1932 and later years, for $M_L>2.5$ in 1975
and later years, for $M_L>2$ for 1992 and later years, and for $M_L>1.5$ 
in 1994 and later years. Since our fits of aftershock decay rate are performed
typically over a time range starting beyond one day 
and extending to several hundred days, the
short-term lack of completeness identified by Kagan [2004] is not an issue here. 
Actually, the problem of the short-term completeness of aftershock catalogues 
is the most acute for large mainshocks [{\it Kagan}, 2004], which allows us to extend the 
study of the seismic decay rate at shorter times for the small mainshock magnitudes.

In order to maximize the size of the data used for our analysis (to improve
the statistical significance) and to test for the stability of our inversion,
we will consider four different sub-catalogs: 
\begin{enumerate}
\item $1932-2003$ for $M_L>3$ ($17,934$ events), 
\item $1975-2003$ for $M_L>2.5$ ($36,614$ events), 
\item $1992-2003$ for $M_L>2$ ($54,990$ events), and 
\item $1994-2003$ for $M_L>1.5$ ($86,228$ events).
\end{enumerate}
The fact that the number of events in these sub-catalogs increases
as their time-span decreases is due to the lowering of the minimum magnitude
of completeness which more than compensates the reducing time interval.

\subsection{Construction of the time series of aftershocks}

\subsubsection{Time-shifted stacked sequences: general principle}

We now describe the method used to determine the validity
of the Omori law and to measure the $p$-value as a function of the mainshock magnitude.

\begin{itemize}
\item We consider all events in a given sub-catalog and discriminate
between mainshocks and triggered events (``aftershocks''). Mainshocks are determined
by using declustering methods which are described below in sections
\ref{1stde} and \ref{2ndde}: to test for the robustness
of our results and their lack of sensivity with respect to the specific choice
of a declustering method, we present two different implementations. 

\item Once the mainshocks
are determined, triggered events are defined as those events
following a mainshock, which belong to a certain space-time neighborhood of it. This
space-time neighborhood is specified below. We thus have
as many triggered (``aftershock'') sequences as there are mainshocks. 

\item In order to test for the predicted dependence of the $p$-value as a function
of magnitude, we need to bin the mainshock magnitudes in intervals $[M_1; M_2]$.
We have chosen the intervals (when available) 
$[1.5; 2]$, $[2; 2.5]$, $[2.5; 3]$, and so on up to $[7; 7.5]$.

\item In each mainshock magnitude interval $[M_1; M_2]$, we consider
all triggered sequences emanating from mainshocks with magnitude in this interval $[M_1; M_2]$.
We translate each triggered sequence so that
their origin of time (their mainshock time) is moved to the common value $t=0$.
Then, we stack all these sequences to obtain a mega-sequence containing all
triggered events coming from mainshocks in the same magnitude interval $[M_1; M_2]$.

\item We then bin the time axis according to a geometrical series, 
and estimate the average rate of events within each bin. The resulting function is fitted 
using the modified Omori law
\be
N(t)= B+ {a \over (t+c)^{p}}~, 
\label{omorilaw}
\ee
where $B$ is a positive parameter introduced to account for the background seismicity
assumed to be superimposed over the genuine triggered sequences. The time shift $c$ 
ensures the regularization of the seismic rate at $t=0$. 
\end{itemize}

Varying $[M_1; M_2]$ allows us to test for a possible dependence of $p$ as
a function of the magnitude of the mainshock. In order to (i)  maximize the size of 
the data sets, (ii) avoid bias due to incompleteness of the catalogs, and
(iii) test the robustness of our conclusions, we will use
\begin{enumerate}
\item for $M_1 \geq 3$: four different sub-catalogs 
($1932-2003$, $1975-2003$, $1992-2003$, $1994-2003$), giving four estimates for the $p$-value;
\item for $[M_1=2.5; M_2=3]$: three sub-catalogs covering
$1975-2003$, $1992-2003$ and $1994-2003$, giving three estimates for the $p$-value;
\item For $[M_1=2; M_2=2.5]$: two sub-catalogs covering
 $1992-2003$ and $1994-2003$, giving two estimates for the $p$-value;
\item For $[M_1=1.5; M_2=2]$: one sub-catalog covering
$1994-2003$, giving a single estimate for the $p$-value.
\end{enumerate}
To explore further for the robustness of our estimates for $p$, we will also impose $c=0$
to test for the sensitivity with respect to the beginning of the stacked time series.
We will also remove bins at random and re-estimate the $p$-values of these pruned
stacked sequences. Putting all these estimates together for a given magnitude
interval $[M_1; M_2]$ allows us to obtain a mean value and a standard deviation.
Note that for each magnitude interval, we have several thousand events in the decay rate function,
ensuring an adequate estimation of the $p$-value. 
For large magnitudes, there are only 
a few mainshocks contributing to the stacked sequence, but each of them have numerous 
aftershocks. For small magnitudes, each mainshock contributes few triggered events (some
have none) but, due to their large number, the stacked sequences have also a sufficiently
large number of events, even for the smallest magnitude intervals.

We now present the two declustering methods that we have implemented to 
select the mainshocks.

\subsubsection{1st declustering technique \label{1stde}}

The first method is essentially the same as defined in
[{\it Helmstetter}, 2003]. First, every event in the catalog is defined as a 
mainshock if it has not been preceded by an event with larger magnitude 
within a fixed space-time
window $T \times d$, with $T=1$ year and $d=50$ km. Looking for events
triggered by this mainshock, we define another space-time window following
it. The time dimension of the window is also set
to $1$ year, whereas the space dimension depends on the rupture length
of the main event. This spatial window is chosen as a circle of radius
equal to the mainshock rupture length 
\be
L=10^{-2.57+0.6M_L}~,
\label{mgmd}
\ee
which is
the average relationship
between $L$ and magnitude $M_L$ for California [{\it Wells and Coppersmith}, 1994].
If any event falls within this space-time window, it is considered as
triggered by the main event. Note that, since the reported uncertainty of
events localization is about $5km$, we set the size of the spatial window to
$5km$ if $L<5km$. We have also checked the stability of the results by
considering a spatial neighborhood of radius $2L$ rather than $L$ for the 
triggered events.

\subsubsection{Second declustering technique \label{2ndde}}

The second declustering technique is the same as the first one, except for
one element: the space window used for qualifying a mainshock is not fixed
to $d=50km$ but is chosen to adapt to the size of the rupture lengths $L(M_i)$ 
given by (\ref{mgmd}) of all
events of all possible magnitudes $M_L(i)$ preceding this potential mainshock. 
In other words, a potential mainshock is selected if there are not preceding
events within one year in its past, which are at a distance less than twice
their own rupture length.
The space-time window for the selection of triggered events is the same
as for the first declustering method.

This second procedure of declustering
is perhaps more natural than the first one, especially for small 
magnitude events. To take an extreme example, in the first method,
an event of magnitude $1.5$ can not be selected
as a main event if it falls within $50 km$ of a previous larger event, even if the
rupture length of the latter is much smaller than $50km$. This
obviously prevents almost all events of magnitude $1.5$ to be considered as 
mainshocks, and thus biases the statistics. Mainshocks of low magnitudes which are
selected by the first declustering method 
will on average occur within areas of very low seismicity rate, so that the composite
triggered sequence will be under-representative.
Moreover, all events located at say
$51km$ of an event of magnitude $M_L=7.5$ are defined as main events, whereas
they probably belong to its triggered sequence. We can thus expect the first
declustering technique to penalize heavily the quality of the stacked sequence
associated with mainshocks of low magnitudes. The second declustering technique do not
show such biases.

\subsection{Results}

\subsubsection{Least-square fit of the seismic 
rate as a function of time with equation (\ref{omorilaw})}
 
\begin{figure}
\begin{center}
\psfig{file=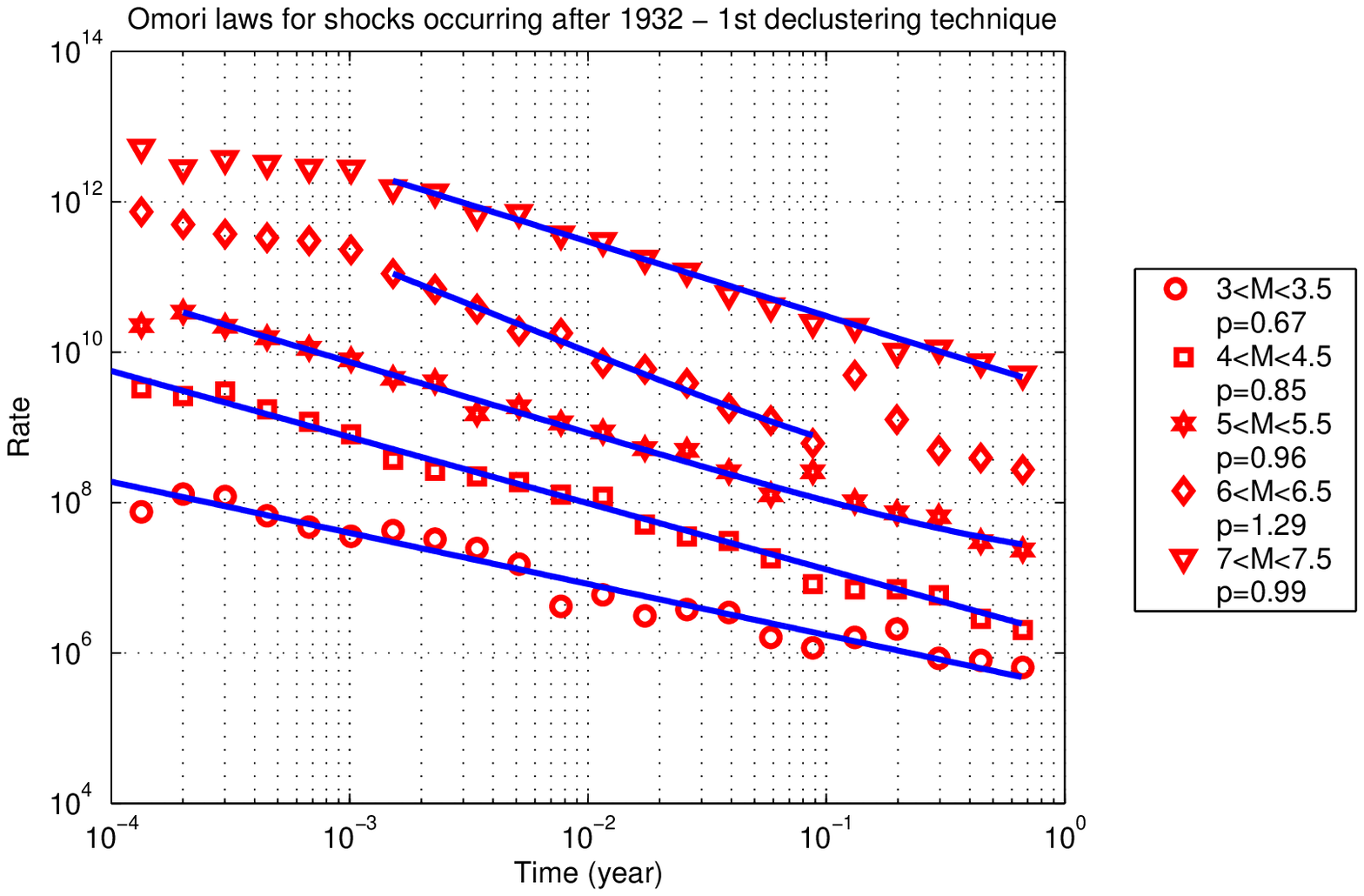,width=14cm}
\caption{\label{omoriplots1} Seismic decay rates of stacked
sequences for several magnitude intervals of the mainshocks,
for the period from 1932 to 2003 when using
the first declustering technique.
}
\end{center}
\end{figure}

\begin{figure}
\begin{center}
\psfig{file=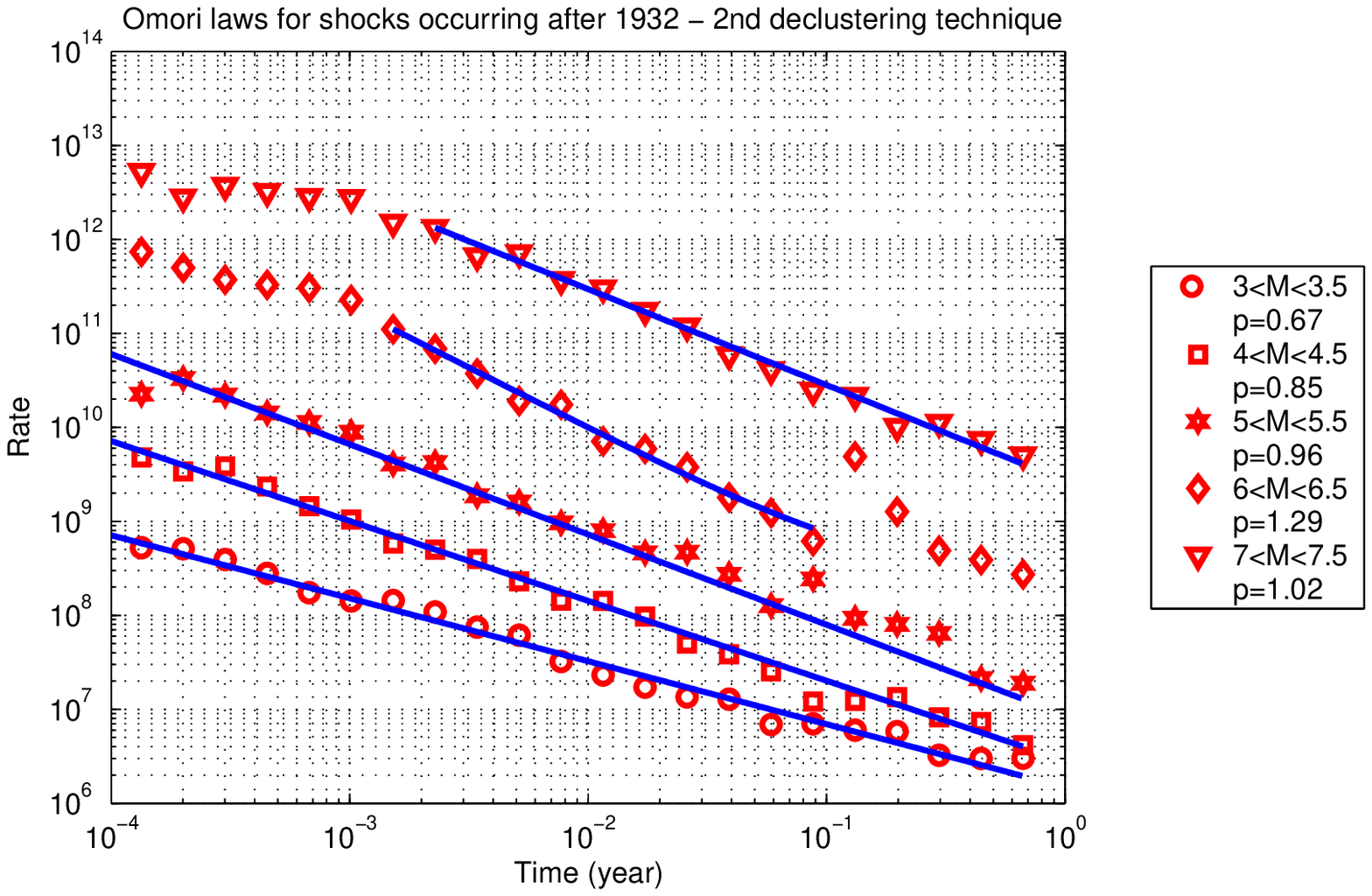,width=14cm}
\caption{\label{omoriplots2} Seismic decay rates of stacked
sequences for several magnitude intervals of the mainshocks,
for the period from 1932 to 2003 when using
the second declustering technique.
}
\end{center}
\end{figure}

\begin{figure}
\begin{center}
\psfig{file=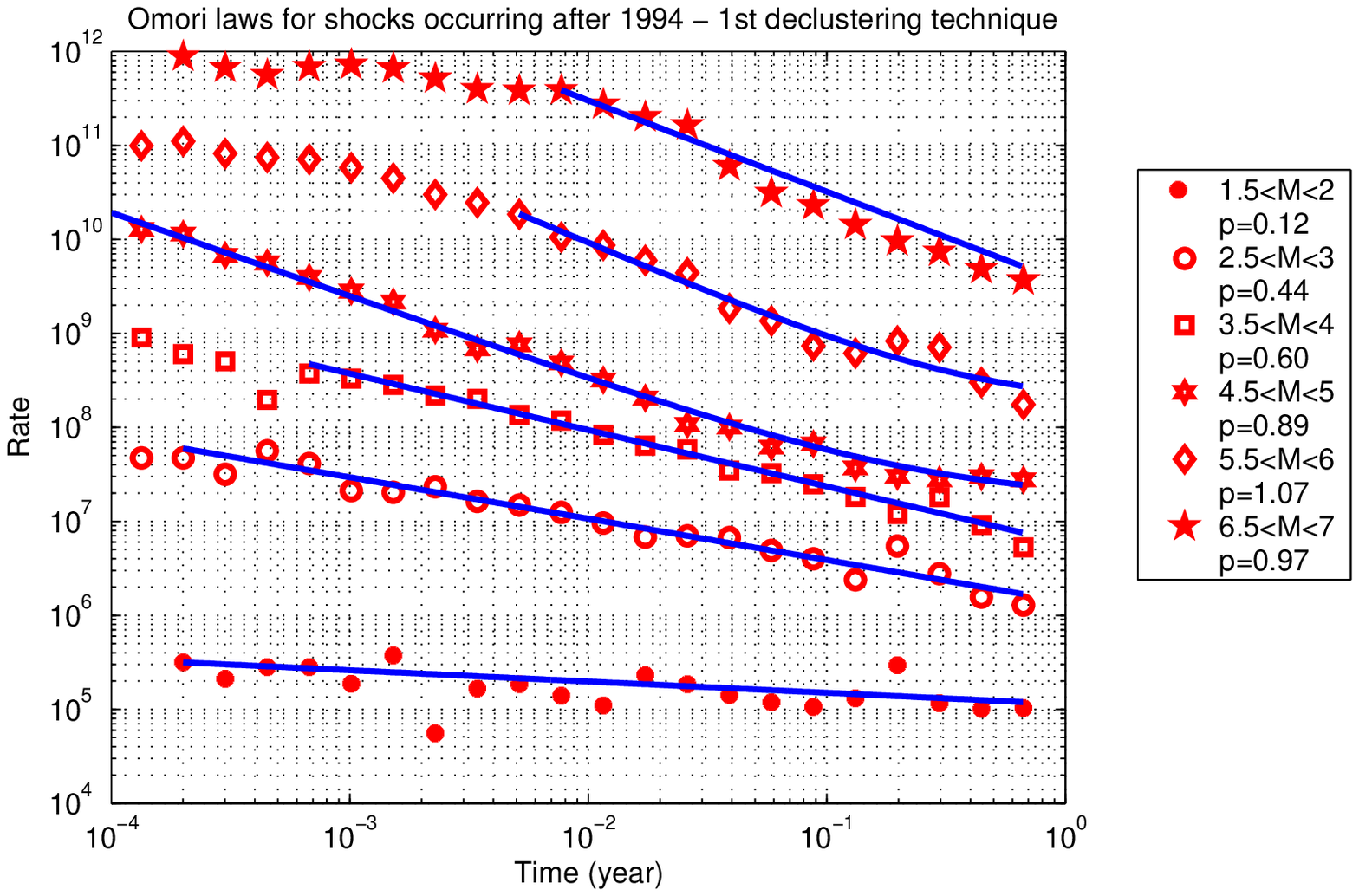,width=14cm}
\caption{\label{omoriplots3} Seismic decay rates of stacked
sequences for several magnitude intervals of the mainshocks,
for the period from 1994 to 2003 when using
the first declustering technique.
}
\end{center}
\end{figure}

\begin{figure}
\begin{center}
\psfig{file=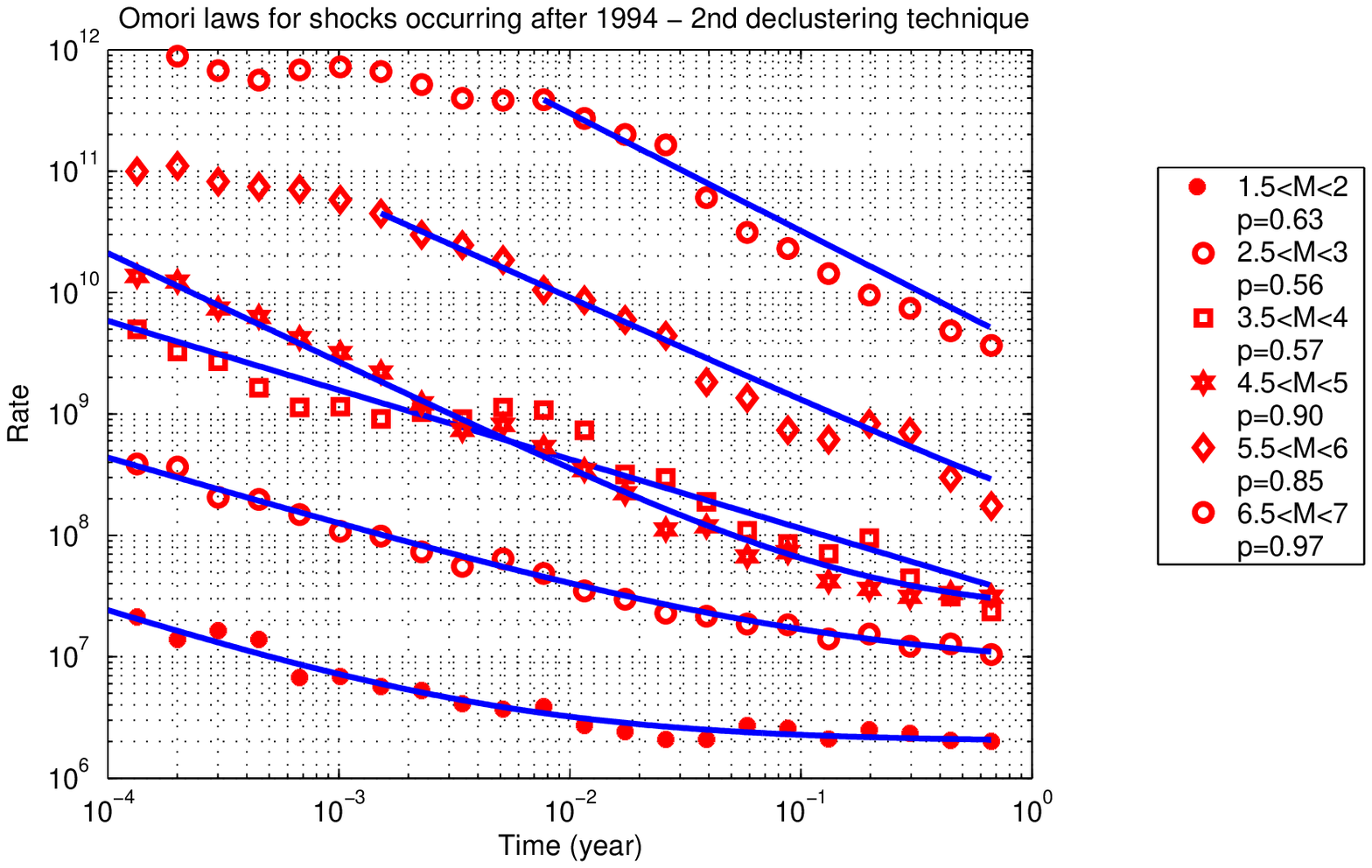,width=14cm}
\caption{\label{omoriplots4} Seismic decay rates of stacked
sequences for several magnitude intervals of the mainshocks,
for the period from 1994 to 2003 when using
the second declustering technique.
}
\end{center}
\end{figure}

Figures \ref{omoriplots1}-\ref{omoriplots4} 
show sets of typical seismic decay rates of stacked
sequences for several magnitude intervals of the mainshocks.
Only half of the individual Omori curves are drawn for the sake of clarity,
as in all similar figures in the remaining of the paper.
Figure \ref{omoriplots1} (respectively Figure \ref{omoriplots2})
 corresponds to the period from 1932 to 2003 when using
the first (respectively second) declustering technique, with mainshock
magnitudes above $M_L=3$.
Figure \ref{omoriplots3} (respectively Figure \ref{omoriplots4})
 corresponds to the period from 1994 to 2003 when using
the first (respectively second) declustering technique,
with mainshock magnitudes above $M_L=1.5$.
Very similar plots are obtained for different time periods and by varying the size from
$L$ to $2L$ of the spatial domain over which the triggered sequences are selected.
For large mainshock magnitudes, the roll-off at small times is due to the observational 
saturation and short-time lack of completeness of triggered sequences [{\it Kagan}, 2004].
In some cases, one can also observe the cross-over to the constant background at large times.

\begin{figure}
\begin{center}
\psfig{file=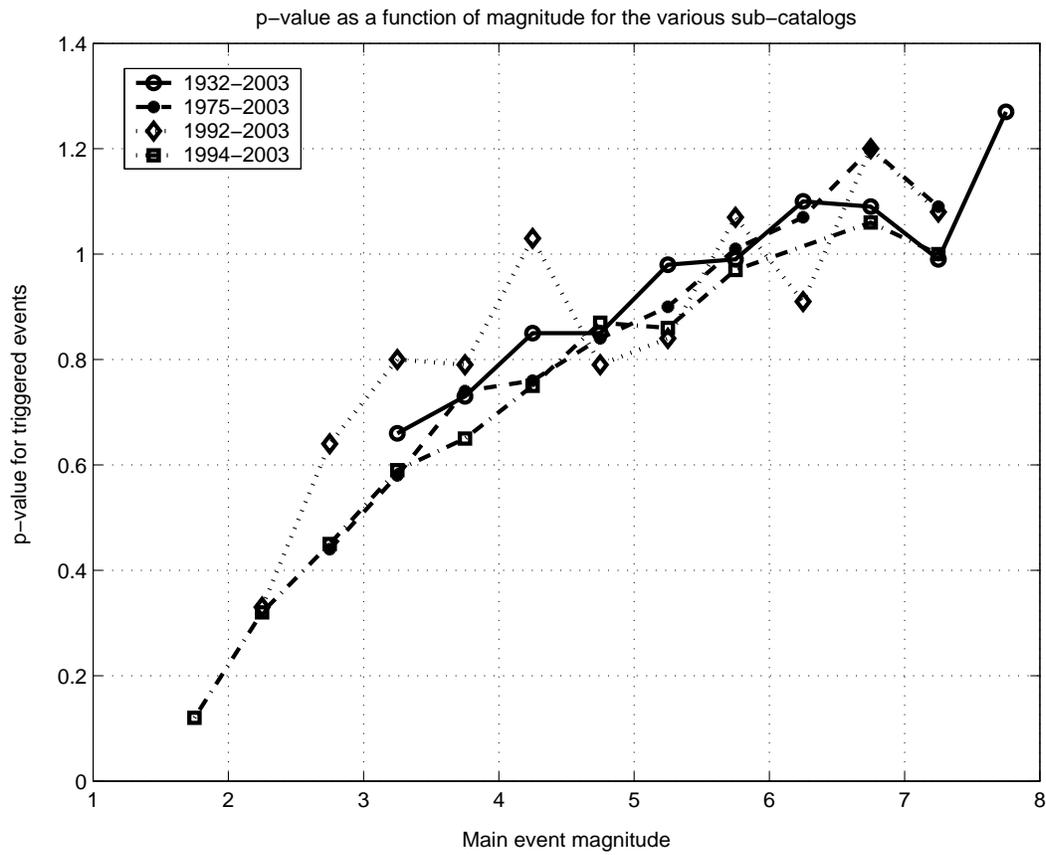,width=14cm}
\caption{\label{F3} $p$-values of the Omori law (\ref{omorilaw}) 
obtained by the procedure described in the text
for mainshocks (defined using the first declustering algorithm)
as a function of the main events' magnitude, for the different
sub-catalogs of lifespans given in the inset.
}
\end{center}
\end{figure}

Figure \ref{F3} shows the fitted
$p$-values as a function of the magnitude of the mainshocks for each of the four
sub-catalogs (the abscissa corresponds to the magnitude $(M_1+M_2)/2$ in the middle of the range
$[M_1;M_2]$ for each interval). We use a standard least-square fit of the seismic 
rate as a function of time with a weight proportional to $t$
for each bin to balance their relative importance. We take into account the possible 
presence of a background term as shown in equation (\ref{omorilaw}).

\begin{figure}
\begin{center}
\psfig{file=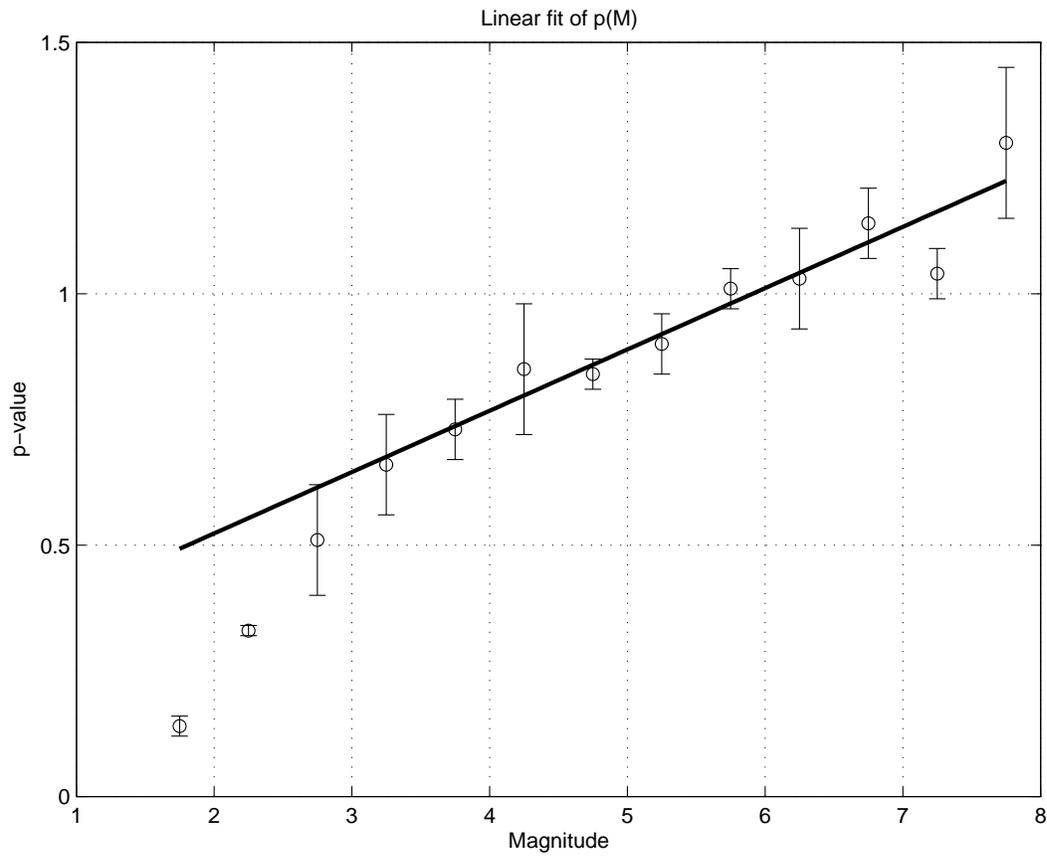,width=14cm}
\caption{\label{F4}  Average $p$-values and error bars obtained from 
Figure \ref{F3} as described in the text. The straight line is the 
linear fit with  $p(M)=0.12M_L+0.28$.
}
\end{center}
\end{figure}

Figure \ref{F4} plots the average $p$-values and their error bars as defined above.
Both figures exhibit a very clear increase of the $p$-value with the mainshock magnitude.
For magnitudes $M_L \geq 3$, a linear fit 
\be
p(M_L) = 0.12 M_L + 0.28
\label{gnmgd}
\ee
is shown as the straight line in Figure \ref{F4}.
The deviation from this linear behavior for the lowest magnitudes $M_L \leq 2.5$
can be attributed to the biases associated with the first declustering technique, as explained above,
when comparing with the results of the second declustering method.

\begin{figure}
\begin{center}
\psfig{file=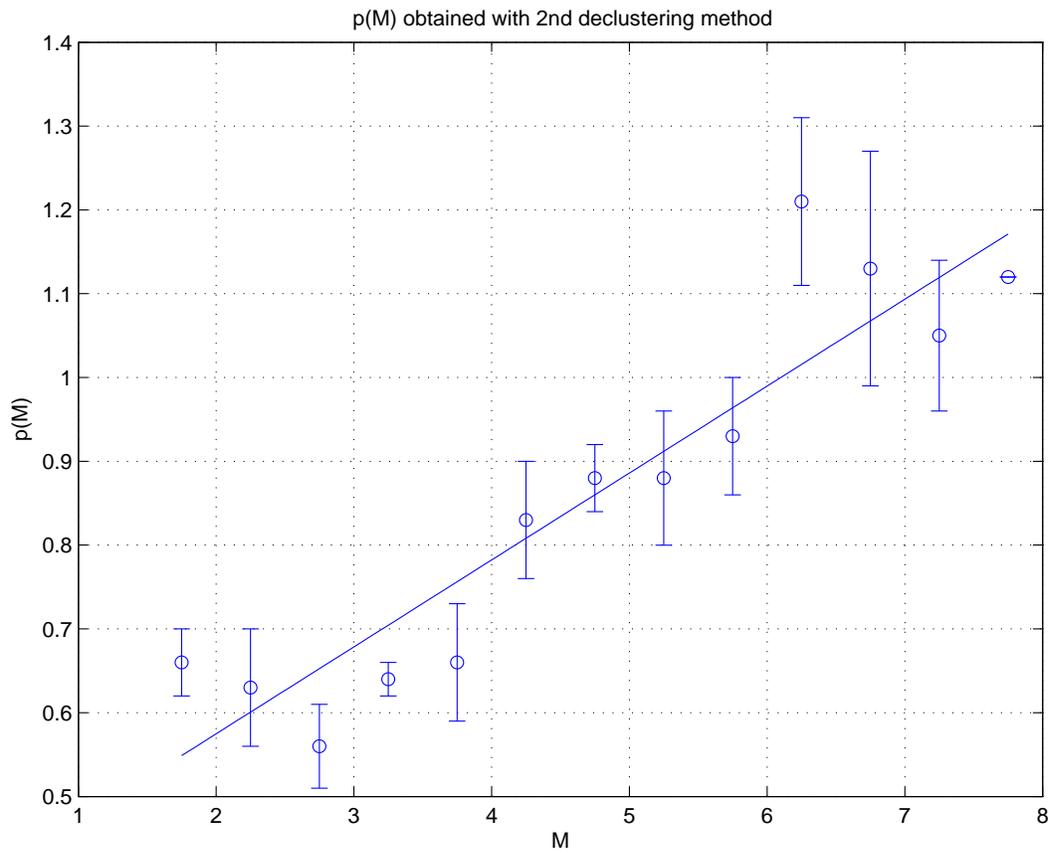,width=14cm}
\caption{\label{F5}  Average $p$-values and error bars obtained 
as described in the text using the second declustering technique.
The straight line is the 
linear fit with $p(M)=0.10 M+ 0.37$.
}
\end{center}
\end{figure}

Figure \ref{F5} shows the $p$-value and its standard
deviation as a function of mainshock magnitude obtained
with the second declustering method. As with the first declustering method, this data
is well-fitted by a linear relationship
\be
p(M_L) = 0.10 M + 0.37~.
\label{gmmgsas}
\ee

\subsection{Maximum Likelihood Estimation method}

In order to check the reliability of the $p$-values obtained in the previous
analysis, we also used a maximum likelihood estimation (MLE). We obtain the MLE
of the $p$-value in a finite time window
from $t$ to $t_U$, by maximizing the probability that the particular
temporal distribution of the $N$ events in that window results from
Omori's law with this particular $p$ value. The most likely $p$ value
is then given by the implicit equation [{\it Huang et al.}, 2000]
\be
{1 \over p-1} + {t_U^{p-1} \ln t - t^{p-1} \ln t_U \over t_U^{p-1} - t^{p-1}} =
 \langle \ln t \rangle_N ~,  \label{hajakk}
\ee
where the $t_n$ are the time occurrences of the $N$ events between $t$
and $t_U$ and
\be
\langle \ln t \rangle_N \equiv {1 \over N} \sum_{n=1}^N \ln t_n ~.
\ee
We fixed $t_U=1$ year, while $t$ was varied
continuously from $10^{-4}$ to $0.5$ year and we solve for $p$
in the implicit equation (\ref{hajakk}). We could thus check both the
value and the stability of $p$ with sample size and to time boundary effects.

\begin{figure}
\begin{center}
\psfig{file=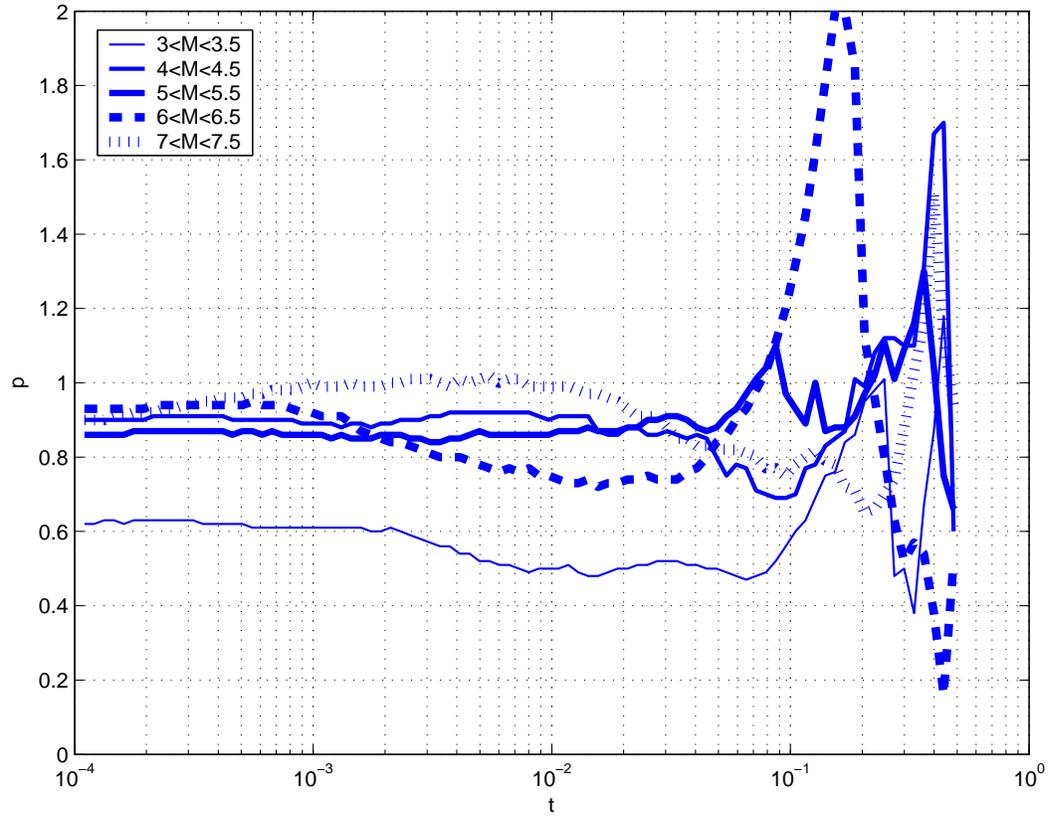,width=14cm} 
\caption{\label{aki_hill_1932_dec_1} Maximum Likelihood Estimation
of the $p$-value (formula (\ref{hajakk}) with $t_U=1$ year, while $t$ was varied
continuously from $10^{-4}$ to $0.5$ year) as a function of $t$
for the post-$1932$ sub-catalog, using the first declustering technique, with $R=2L$, for
different magnitude ranges.}
\end{center}
\end{figure}

Figure \ref{aki_hill_1932_dec_1} shows $p$ as a function of $t$ for the post-$1932$
sub-catalog, using the first declustering technique, with $R=2L$, for
different magnitude ranges. This figure should be compared with Figure
\ref{omoriplots1}. One can observe that, for $3<M<3.5$, $p$ varies very slightly
with time $t$, with a value of about $0.6$, whereas the least-square fit gave
$p=0.67$ for the whole time range (corresponding here to $t=10^{-4}$
year). For $4<M<4.5$, $p$ is close to $0.9$, while the least-square fit gave
$p=0.85$. The case $5<M<5.5$ is different as the maximum
likelihood yields a $p$-value of about $0.85$ whereas the least-square fit
gave $p=0.96$. It should be noted here that the least-square fit detected a significant
non-zero background value $B$, and that the MLE method
doesn't take into account such a term, which leads to an underestimation
of $p$ as the time-distribution is flatter at larger times (which is
also expressed by a strong drop of $p$ at large time $t$). The case
$6<M<6.5$ yields $p$ close to $0.8$ whereas the least-square fit gave
$p=1.29$, but the MLE method also takes account of events
for times larger than $0.1$, where strong secondary aftershock
activity disturbs the distribution and would tend to define a power-law
with lower exponent. Considering the $7<M<7.5$ range, where no such bias
occurs, the MLE method gives $p$ very close
to $1$ for $10^{-3}<t<10^{-2}$ whereas $p=0.99$ for the least-square fit. The
conclusion is thus that, when a background $B$ is absent, $p$-values
obtained with both methods agree very well, which strenghtens our belief
in the reliability of the exponents determined with the least-square fit. This
least-square fit has here the advantage that it takes into account the
possible existence of a non-zero background term $B$.

\begin{figure}
\begin{center}
\psfig{file=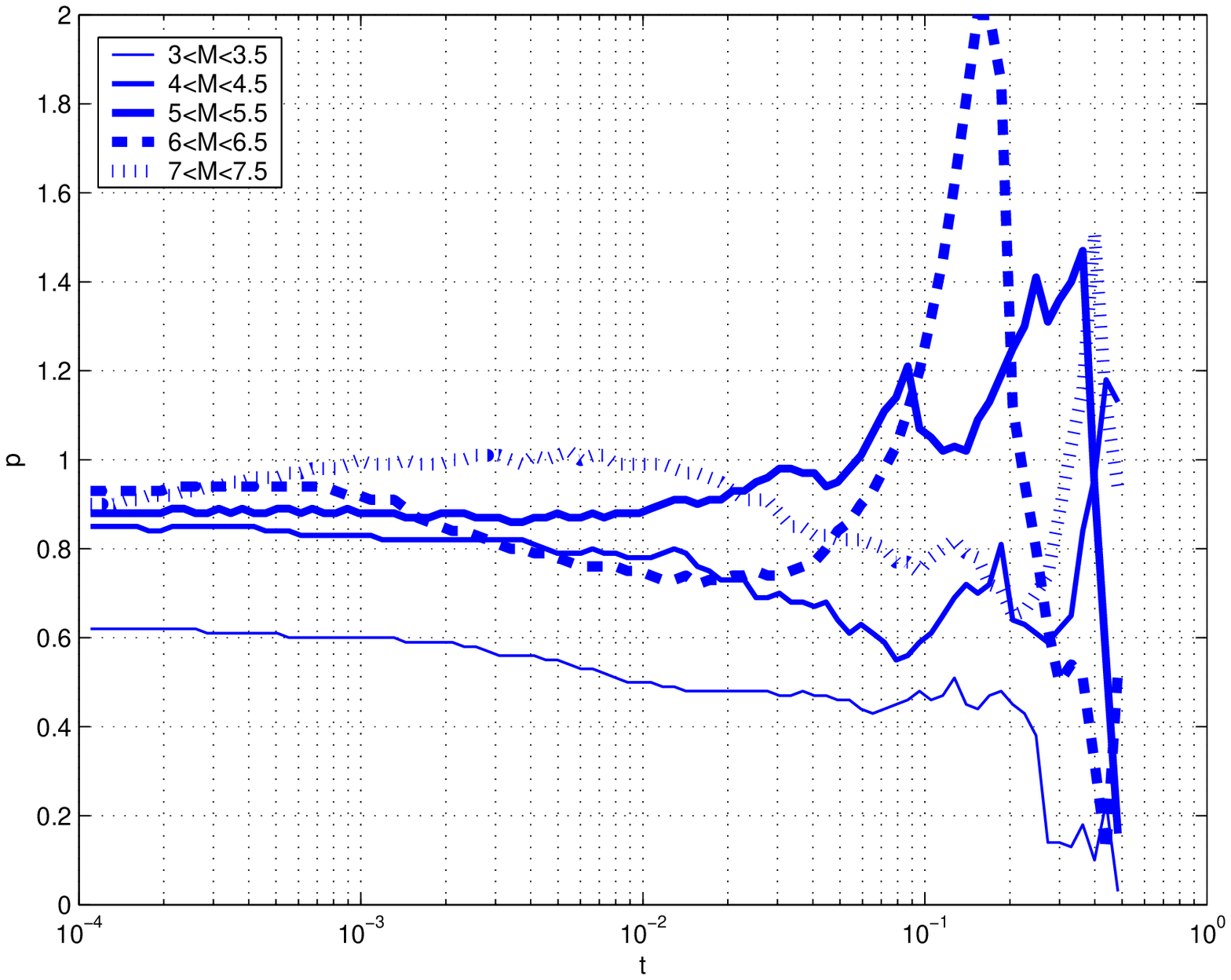,width=14cm} 
\caption{\label{aki_hill_1932_dec_2} Same as Figure \ref{aki_hill_1932_dec_1}
for the post-$1932$ sub-catalog, using the second declustering technique, with $R=2L$, for
different magnitude ranges. 
}
\end{center}
\end{figure}

Using the same data set and the $2nd$
declustering method, we obtained Figure \ref{aki_hill_1932_dec_2}, which
can be compared to Figure \ref{omoriplots2}. Once again, exponents agree very well
(except for the $6<M<6.5$ range, for reasons explained above).

\begin{figure}
\begin{center}
\psfig{file=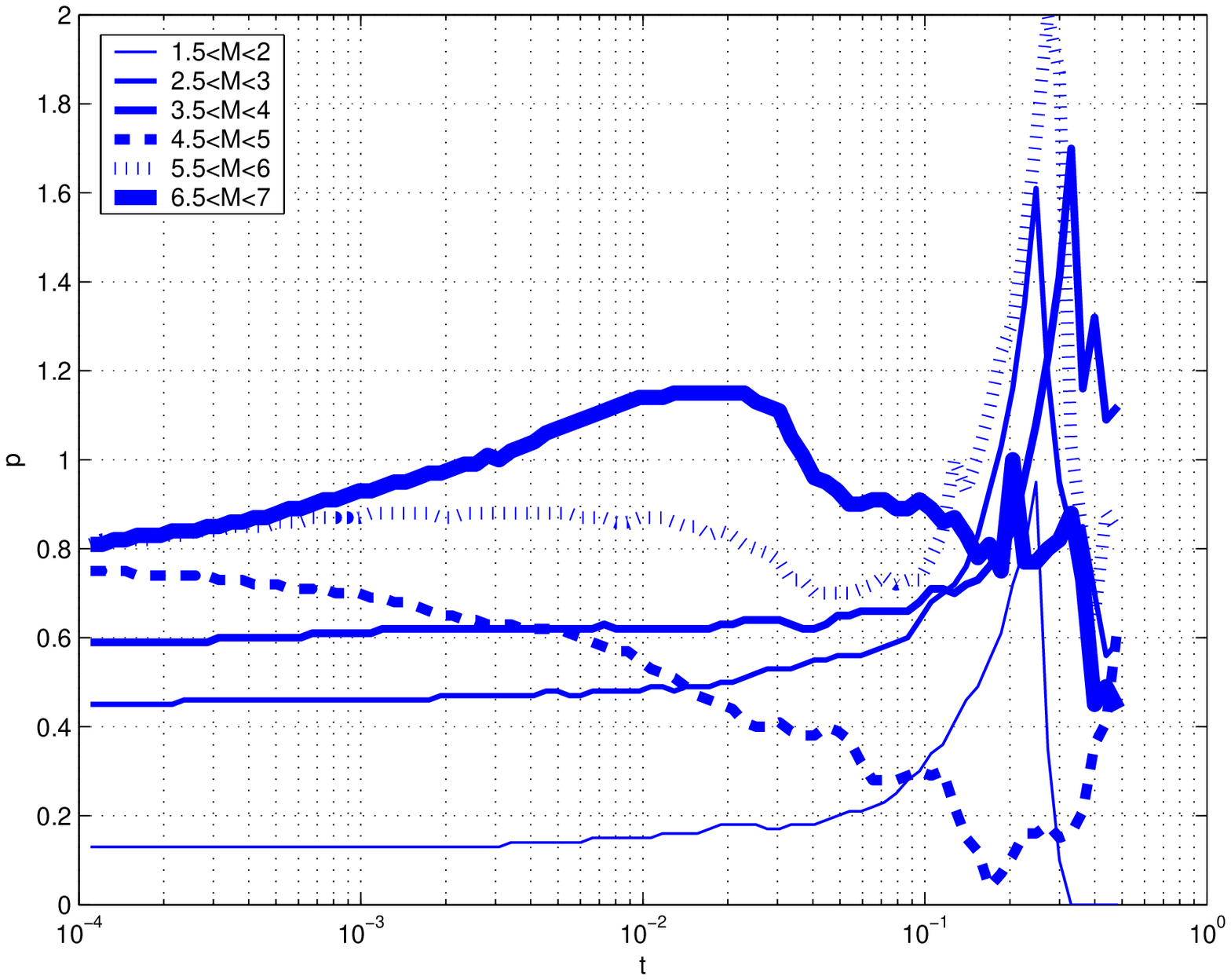,width=14cm} 
\caption{\label{aki_hill_1994_dec_1} Same as Figure \ref{aki_hill_1932_dec_1}
for the post-$1994$ sub-catalog, using the first declustering technique, with $R=2L$, for
different magnitude ranges. 
}
\end{center}
\end{figure}

We also check the consistency of both methods for the post-$1994$
catalog. The MLE provided us Figure
\ref{aki_hill_1994_dec_1} for the first declustering method, which has to
be compared with Figure \ref{omoriplots3}. We note that $p$-values perfectly agree up
to range $3.5<M<4$. For the next magnitude range, the
MLE provides a $p$-value that is continuously
decreasing with $t$. This can be rationalized by a look at the least-square
fit, which shows that the distribution converges toward a constant
background rate at large times, so that the ML inversion
is systematically biased towards low values. The $5.5<M<6$ gives $p$
close to $0.9$ compared with $1.07$ with the least-square fit. Once again, the
existence of non-zero background term certainly biases the MLE.
The last magnitude range, $6.5<M<7$ offers the
largest discrepancy between $p$-values, but it should be noted that for
times larger than $0.01$ year, the $p$-value varies between $1.15$ and
$0.8$, so that the average maximum likelihood value agrees well with the
least-square inversion. Even in the worst case ($t=10^{-2}$ year), the MLE
method even emphasises the variation of $p$ with magnitude.

\begin{figure}
\begin{center}
\psfig{file=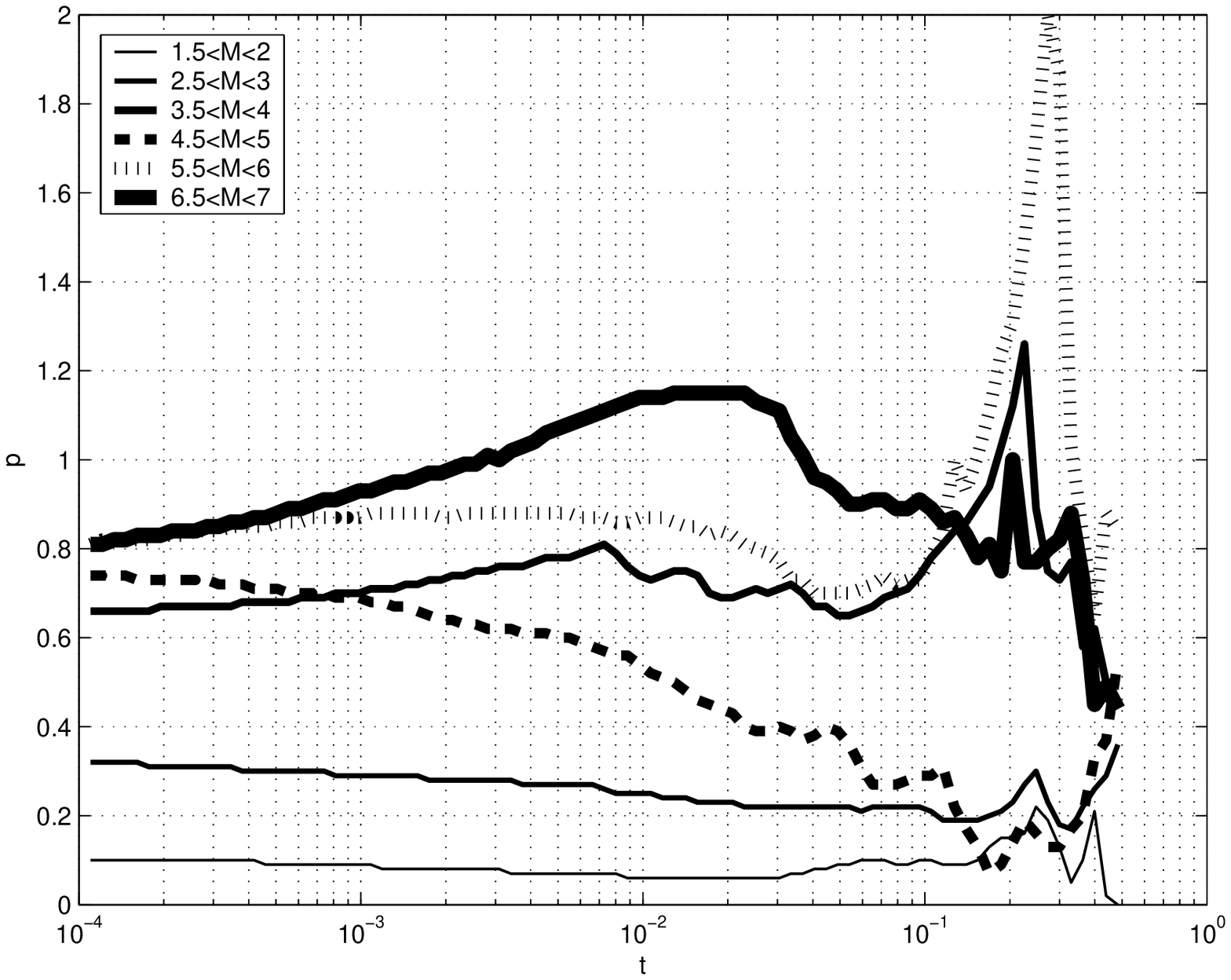,width=14cm} 
\caption{\label{aki_hill_1994_dec_2} Same as Figure \ref{aki_hill_1932_dec_1}
for the post-$1994$ sub-catalog, using the second declustering technique, with $R=2L$, for
different magnitude ranges. 
}
\end{center}
\end{figure}

The results for the $2nd$ declustering technique on the same sub-catalog are
displayed in Figure \ref{aki_hill_1994_dec_2} (which should be
compared with Figure \ref{omoriplots4}). The $p$-value for the lowest magnitude ranges
are quite low ($0.2$-$0.3$) but this is once again due to the background
rate. The value obtained in the $3.5<M<4$ range ($p$ around $0.7$) is
strongly oscillating. This is indeed also the case for the time
distribution shown in Figure \ref{omoriplots4}, for which $p=0.57$. Note that the
MLE method yields $p=0.65$ for $t=10^{-4}$, close to the
least-square fit value obtained on the same data range. We have the same comments 
as above for the $4.5<M<5$ which displays a noticeable background term.
The next magnitude range gives a $p$-value a bit less than
$0.9$, whereas the least-square fit gave $p=0.85$. The same comments as above
apply for the $6.5<M<7$.

Overall, the conclusion is that, when the background rate is absent from the
seismic decay, $p$-values inverted from both methods agree very
well, and that even when we observe a discrepancy, the ML 
inversion amplifies the variation of $p$ with magnitude.

\subsection{Tests of our procedure with synthetic ETAS catalogs}

To test the reliability and robustness 
of our results, we now analyze simulated catalogs 
with known statistical properties following exactly the same
procedure as for the real catalogs. We 
generated synthetic catalogs using the ETAS model, running
the code of K. Feltzer and Y. Gu (spring 2001) modified by A. Helmstetter (2003). 
This model has a fixed magnitude-independent Omori $p$-value as an input.
Thus, by construction, synthetic catalogs generated with the ETAS model
should exhibit Omori laws with magnitude-independent exponents. Applying
our procedure to such synthetic catalogs, which are known to be very similar
to real catalogs in many of their statistical properties 
[{\it Helmstetter and Sornette}, 2003], allows us
to investigate whether the magnitude-dependence of the 
$p$-value reported above could result
from some bias introduced by our analysis rather than being 
a genuine property of earthquake catalogs.

In the ETAS model, a main event of magnitude $M$ triggers its own
primary aftershocks according to the following distribution in time and space
\begin{equation}
\phi_m (r,t)~ dr~ dt = K~ 10^{\alpha (M-M_0)}
~\frac{\theta~c^{\theta}~dt}{(t+c)^{1+\theta}} ~
\frac{\mu~d^{\mu}~dr}{(r+d)^{1+\mu}}~,
\label{nmgjedl}
\end{equation}
where $r$ is the spatial distance to the main event (considered as a
point process). The spatial regularization distance $d$ accounts for
the finite rupture size. The power law kernel in space with
exponent $\mu$ (this exponent should not be confused with that
of the distribution of the stress fluctuations defined in (\ref{vcmnfdlqa}))
quantifies the fact that
the distribution of distances between pairs of events is
well described by a power-law [{\it Kagan and Jackson}, 1998]. 
In addition, the magnitude
of these primary aftershocks is assumed to be distributed according
to the Gutenberg-Richter law of parameter $b$. The ETAS model assumes that
each primary aftershock may trigger its own
aftershocks (secondary events) according to the same law, the secondary
aftershocks themselves may trigger tertiary aftershocks and so on,
creating a cascade process. The exponent $1+\theta$ is not the observable
Omori exponent $p$ but defines the local (or direct) Omori law
[{\it Sornette and Sornette}, 1999; {\it Helmstter and Sornette}, 2002a].
The two exponents $b$ and $\theta$ should not be confused with 
those used for the Green function of viscous relaxation.

Using the ETAS code, we thus generated a catalog of earthquakes
located within a three-dimensional slab of horizontal dimension
$500 \times 500$ km$^2$ and thickness $20$ km.
We added a noise with amplitude of
$5$ km to the position of each event to simulate the location uncertainty of real catalogs.
The parameters of the ETAS model were chosen as follows:
$b=0.9$, $\theta=1.1$, 
$c=10^{-5}$ day, $K=0.002$, and $\alpha=b=0.9$.
The characteristic spatial distance $d$ in the ETAS kernel 
was taken equal to the event rupture length (which we deduced from its magnitude,
according to the {\it Wells and Coppersmith (1994)} relationship (\ref{mgmd})), 
while the spatial decay
exponent $\mu$ was fixed to $1.0$.
The minimum magnitude of generated events was set to $M_0=0.5$,
and we introduced a truncation of the Gutenberg-Richter distribution such that
no event of magnitude $M_l>8.0$ are allowed. The rate of background events
was set to $50$ events/day. The obtained synthetic catalog
is $25$ years long, and only the $10$ last years are retained to
minimize temporal edge effects at the beginning of the time-series.
Finally, only events of magnitude larger
than $1.5$ were kept in the catalog, to mimic the effect of a magnitude 
detection threshold of real seismic networks. 
We performed several simulations until we generated a catalog similar 
to the post-$1994$ sub-catalog we previously analyzed: by similar, we
mean that the synthetic catalog has approximately the same total number
of events and the same number of events with magnitude $M_l>7.0$.
With these parameters, the branching ratio 
(mean number of triggered events per shock) is $n=0.983$ and the characteristic time
$t^*$ below which the bare Omori law $\sim 1/t^{1+\theta}$ is renormalized into 
the observable global Omori law $\sim 1/t^{1-\theta}$ is
$t^{*}=1.1 \cdot 10^{10}$ years. Thus, the $p$-value of any aftershock
sequence is by construction equal to $p=1-\theta=0.9$ in the time interval 
$[10^{-4} {\rm year} ; 1 {\rm year}]$
that we used for our analysis, which is the same time scale used for our analysis of aftershock
sequences of real catalogs. 

\begin{figure}
\begin{center}
\psfig{file=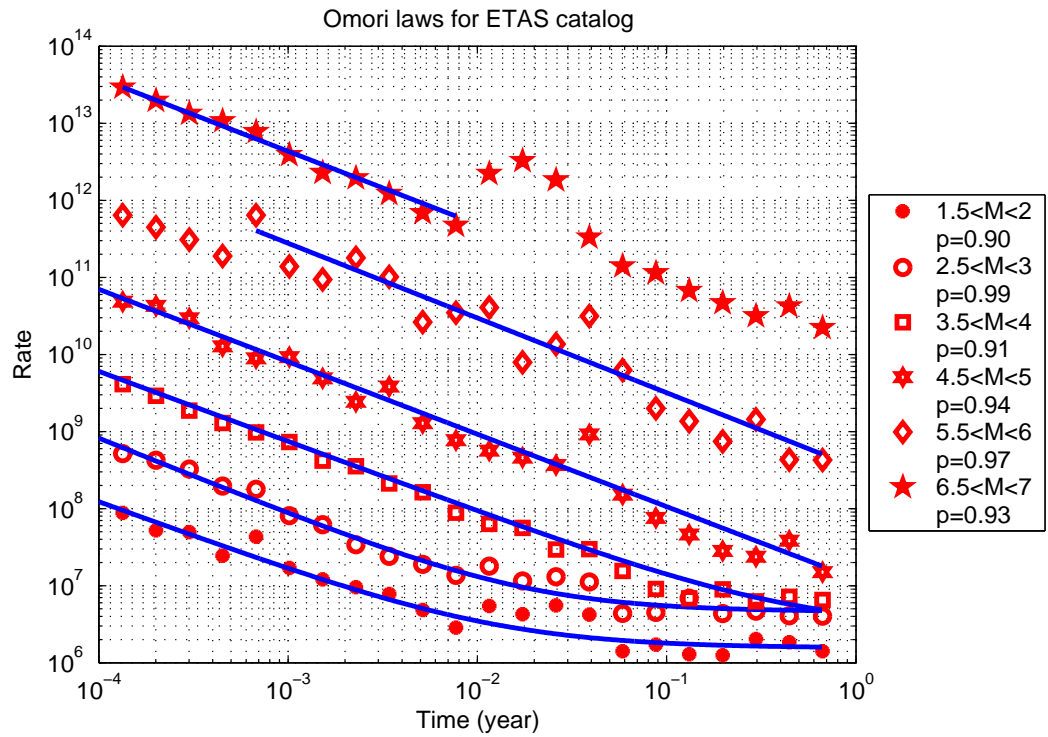,width=14cm} 
\caption{\label{etas_fits} Same as Figure \ref{omoriplots2} for the 
synthetic catalog generated with the 3D ETAS model,
using the second declustering technique, with $R=2L$, for
different magnitude ranges. 
}
\end{center}
\end{figure}

The selection of mainshocks and aftershocks in our synthetic catalog
was performed using the second declustering algorithm
(with $R=2L$) described in section \ref{2ndde}, and we used the same 
stacking method to derive empirically
the $p(M_l)$ relationship. The results obtained with the fits 
of the logarithm of the stacked seismic rates as a function of the logarithm 
of time are shown in Figure \ref{etas_fits}. 
The $p$-values are found independent of the magnitude of the mainshock and close to 
the correct value. There is actually a tendency for the $p$-value to decrease
with the mainshock magnitude (which is the opposite of the effect predicted by our model
and reported for the real catalogs). The origin of this effect is the following:
for mainshocks of small magnitudes,
there are so many stacked time-series that fluctuations average out allowing
to retrieve a precise $p$-value; in constrast, after mainshocks of large magnitudes, 
strong secondary aftershock sequences often occur which introduce large fluctuations.
As there are only a few stacked aftershock series associated with the relatively rare large mainshocks,
the fluctuations of the average seismic rates after the large mainshocks
do not average out; bursts of strong aftershocks tend to bias the $p$ downward
leading to its under-estimation. To minimize this effect for events of large magnitudes, we
fitted the Omori law over the beginning part of the time series and removed
the end of the time-series to compute $p$ (see Figure \ref{etas_fits}). 

\begin{figure}
\begin{center}
\psfig{file=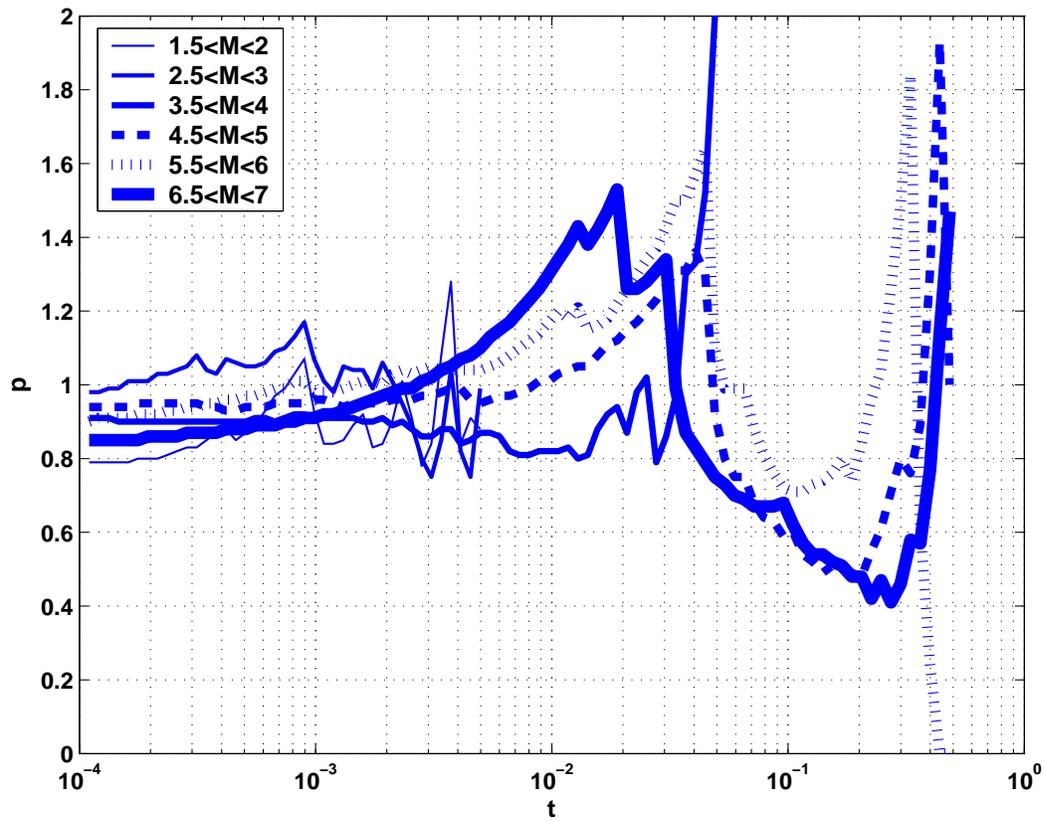,width=14cm} 
\caption{\label{etas_maxlik} Same as Figure \ref{aki_hill_1932_dec_1}
for the synthetic catalog generated with the 3D ETAS model. The MLE formula
(\ref{hajakk}) is applied in a finite time window
from $t$ to $t_U$, following the same method as for the real catalogs. 
The upper value $t_U$ is $0.01$ year for magnitudes up to $3$, $0.1$ year for the
$3.5-4$ magnitude range, and $1$ year for larger magnitude ranges, so as
to minimize bias due to the background seismicity.
}
\end{center}
\end{figure}

We also inverted $p$ on the same synthetic data using the maximum likelihood
formula (\ref{hajakk}) in a finite time window
from $t$ to $t_U$, following the same method as for the real catalogs. We span
$t$ up to $t_U/2$.
Figure \ref{etas_maxlik} plots the $p$-value in different mainshock magnitude ranges
as a function of $t$. Since the MLE method is sensitive to the background seismicity, 
using the lessons from our previous analysis on the real catalogs.
we restricted $t_U$ to avoid biases due to the background
seismicity. Thus, for magnitudes up to $3$, we did not take into account data for
times larger than $0.01$ year. This upper cut-off has been set to $0.1$ year for the
$3.5-4$ magnitude range, and no cut-off (that is, $t_U = 1$ year) 
was imposed for larger magnitude ranges
since their larger triggered seismicity makes the background  negligible up to 
one year after the mainshocks. The results are not sensitive to these specific
cut-offs as long as they correspond to negligible background seismicities.
Figure \ref{etas_maxlik} clearly shows that the $p$-values cluster around $p=0.9$ whatever
the magnitude range. Oscillations occurring at large times correlate
with oscillations observed on the binned data, and betray the influence of secondary
aftershock sequences.

These results and the accuracy of the recovered $p$-values in our synthetic
catalog strengthen one's confidence in our reported results that the magnitude
dependence of the $p$-value in the real Southern California catalogs 
is a genuine effect and not an
artifact of our data analyzing procedure.

\section{Discussion}

\subsection{Summary}

We have proposed a new physically-based ``multifractal stress activation''
model of earthquake interaction and triggering
based on two simple ingredients: (i) seismic ruptures result from activated processes
giving an exponential dependence of the seismic rate 
on the local stress; (ii) the stress relaxation has
a long memory, typically larger than one year. 
The combination of these two effects gives rise in a rather general way
to seismic decay rates following mainshocks that can be well-described by 
apparent Omori laws with exponents $p$ which are linearly increasing with the 
magnitude $M_L$ of the mainshock. This $p(M_L)$ dependence can be interpreted
as a temporal multifractality, that is, as a continuous spectrum of exponents, each exponent being
associated with a given singularity strength (mainshock magnitude). 

While rather general, these predictions require, within our model, that 
the two exponents $\theta$ (stress relaxation) and $\mu$ (stress strength distribution)
verify approximately the condition $\mu (1+\theta) = 1$. We stress that
the special case $(\mu=2, \theta=-1/2)$, which has been shown to give an exact
multifractal process, obeys this condition. Since $\theta$ and $\mu$ are two inputs, 
our theory is mute on the possible origin of this condition. The fact that such a 
constraint appears as the condition to observe an Omori power law is a
very interesting prediction to test in the future. 
As a bonus, this condition predicts the multifractality
expressed by a dependence of the Omori exponent on the earthquake magnitude.
These results can also be seen to provide a generalization
to the multifractal random walk $(\mu=2, \theta=-1/2)$ of [{\it Muzy and Bacry}, 2002;
{\it Schmitt and Marsan}, 2001],
by showing that multifractality is not an isolated property 
at a single point in the plane
$(\mu, \theta)$ but occurs over a line of co-dimension $1$. At this stage,
we can only conjecture that a broader model embodying the self-organization
of the stress field and earthquake space-time organization will lead
to the prediction that indeed the two exponent $\mu$ and $\theta$ are not independent
but are linked by the condition $\mu (1+\theta) = 1$.

These predictions have been tested by a careful and detailed analysis of earthquake
sequences in the Southern California Earthquake catalog.
The robustness of the results obtained with respect to different time intervals, magnitude ranges, 
declustering methods suggests that we have discovered a new important fact of 
seismicity: the apparent power law relaxation of seismic sequences triggered by mainshocks
has indeed an exponent $p$ increasing with the mainshock magnitude by approximately $0.1-0.15$ unit
for each magnitude unit increase. The fits (\ref{gnmgd}) and (\ref{gmmgsas}) of the data
are in agreement with the theoretical prediction (\ref{mgffz}) of the proposed 
multifractal stress activation model.

\subsection{Intuitive ``proof'' that $p(M_L)$ increases with $M_L$ for any multifractal
generalization of Omori's law}

Here, we give an heuristic and intuitive reason why, if $p$ varies with $M_L$, this
can only be by increasing with the mainshock magnitude.

Consider the plate tectonic process which continuously produces earthquakes,
which trigger other earthquakes and so on, with a productivity increasing
with the earthquake magnitude. Let us study the temporel evolution in a fixed 
spatial domain. This temporal evolution can be viewed to 
define a statistically stationary measure defined
on the temporal axis, the measure determining the rate of earthquakes at any 
possible instant. A general method for quantifying such measure is to calculate
its multifractal spectrum. For instance, if the seismic rate is constant,
the measure is uniform and the multifractal spectrum is reduced to a point
of dimension $f(\alpha)=1$ at the singularity strength $\alpha=1$.
Note that this notation $\alpha$ refers in the present discussion to the exponent
of a local singularity and not to the exponent of the productivity law.

An Omori sequence with exponent $p$ corresponds to a singularity (to the right) 
equal to $1-p$ (for $p \neq 1$). Let us calculate the singularity spectrum
of the seismic rate measure. This is usually done via the calculation of 
moments of order $q$, large positive $q$'s corresponding to small $\alpha$'s,
that is, to strong singularities (see for instance Chapter 5.2 in
[{\it Sornette}, 2004]). Now, a large earthquake triggers 
a strong burst of seismicity, giving rise to a strong singularity. From the 
relation $\alpha = 1 - p$, to be consistent with the multifractal description,
a large earthquake must be associated with a strong singularity, a small $\alpha$,
hence a large $p$.
Reciprocally, small moment orders $q$ select weak seismic sequences, which are
thus associated with small local mainshocks. Small $q$'s are associated with large 
$\alpha$'s, hence small $p$'s.  
Thus, any generalization that allows for a dependence of $p$ on the mainshock magnitude
necessarily leads to an exponent $p$ increasing with the magnitude.

By a similar argument in the space domain, the exponent of the spatial 
decay of the seismic rate induced by a mainshock of magnitude $M_L$ should
increase with $M_L$.  Thus, in this view, the ETAS model is nothing but the mono-fractal
approximation of the more general multifractal description of seismicity.

\subsection{Self-similarity of earthquakes}

A central empirical observation in seismology is the unability 
to discriminate between rupture processes of large and small earthquakes. 
The signature of this self-similarity of individual seismic events 
translates into the invariance of the stress drop (inverted from seismic
waves) with rupture size. It would thus seem that Figure \ref{F3} for instance
is contradicting this empirical law since 
the statistics of the time-series of triggered events depend on the mainshock magnitude.
This could be interpreted as reflecting different rupture mechanisms, 
thus breaking scale invariance. However, such an interpretation is incorrect. First, it is
now well-understood that mono-fractality embodied in a universal (Omori) power law
is not the only signature of scale invariance. Since Parisi and Frisch [1985] and
Halsey et al. [1986], several complex systems have been shown 
to exhibit an extended form of scale invariance, called multifractality. The $p(M)$
dependence documented here is an example of such multifractality.
Second, the main ingredients of our theoretical model are scale invariant:
(i) the triggering mechanism of a single
shock is stress activation, independently of the magnitude of the event to be
nucleated; (ii) we make no assumption the stress drop during an event;
(iii) the Cauchy law (\ref{mgbmls}) (exponent $\mu=1$ in (\ref{vcmnfdlqa}))
was suggested by Kagan [1994] using
the self-similarity argument, and our theoretical results show that even in
that case $p$ depends on $M$. Once again, the long memory of the stress 
relaxation kernel coupled
with the exponential stress activation are sufficient to predict such a magnitude
dependence, which is thus compatible with the self-similarity
hypothesis.

\subsection{Alternative models}

Could these observations be interpreted differently than with the 
multifractal stress activation model? For instance, 
a change of the apparent Omori exponent $p$ is consistent with the ETAS 
model if the critical branching
ratio $n$ (average number of triggered events per mainshock) is less than unity: 
if $n$ is close to its critical value $1$, $p$ is close
to $1-\theta$, while for $n<1$, $p$ goes from 
$1-\theta$ at short time scale to $1+\theta$ beyond a characteristic time
scale $t^* \sim (1-n)^{-1/\theta}$ 
[{\it Sornette and Sornette}, 1999; {\it Helmstetter and Sornette}, 2002a]. The parameter
$\theta$ is often found in the range $0.2-0.3$ for large shocks. Thus, to explain our results
(\ref{gnmgd}) and (\ref{gmmgsas}), one would need to invoke 
a larger value for $\theta$ of the order of $0.5$
and to have $n$ depend on mainshock magnitudes. Or if the measurement for 
small mainshock magnitudes are performed at shorter times than for larger
mainshocks, this would lead to an increase of the $p$-value because
short times should be controlled by the exponent $1-\theta$ while longer time 
reveal the exponent $1+\theta$. However, this last explanation is ruled
out by the fact that our determination of the $p$-values is performed
on the same time interval for all magnitudes.

One could also perhaps argue that small 
earthquakes reveal the critical state of the crust corresponding
to $n$ close to $1$ while large earthquakes move the crust away from criticality so that their
triggered aftershock sequences correspond to a small branching ratio $n<1$, hence the 
larger exponent. This picture has been recently advocated under the concept
of ``intermittent criticality'' 
(see for instance [{\it Jaum\'e and Sykes}, 1999; {\it Goltz and Bose}, 2002;
{\it Ben-Zion et al.}, 2003; {\it Bowman and Sammis}, 2004]).
While we cannot exclude that intermittent criticality is the explanation for our 
results, this interpretation has only qualitative predictive power and needs more fine-tuning
than the multifractal stress activation model since the later predicts precisely the 
observed linear dependence of $p(M)$. For the ETAS model to explain our results, we would need
a magnitude dependence of the productivity parameter $\alpha$ which 
does not seem to be observed
[{\it Helmstetter}, 2003] (note however that Helmstetter's measure of $\alpha$
are probably underestimated as they have relied on the assumption
of a constant $p$-value and have used a fixed space domain size
independent of mainshock magnitude to select them). 
In addition, the multifractal stress activation model
is physically-based while the ETAS model is only phenomenological and less attractive
as an ``explanation.''

\subsection{Other predictions of the multifractal stress activation model}

\subsubsection{Temperature dependence of $p$-values}

Another prediction of the multifractal stress activation model is the 
linear dependence of the $p$-value $\propto \beta$ in 
(\ref{mgffz}) as a function of the inverse of the temperature. This implies that
the strong dependence of the $p$-value as a function of mainshock magnitude
should be more visible with a larger amplitude for cold regions. This suggests
a tantalizing new interpretation of the correlations reported 
between the $p$-value and thermal flux [{\it Kisslinger and Jones}, 1991].
These authors observed a positive 
correlation, superficially in line with the intuitive idea that hotter material
relaxes stress at larger rates. We must point out, however, that they did not
take account of uncertainties on measured heat flow, and that
such a correlation was obtained by inverting
$p$ on a few dozens of individual aftershock sequences following mainshocks
of different magnitudes (from $5.1$ to $7.5$).
Using a world-wide catalog of events of magnitudes larger then $5$,
within which they selected nine subregions, Marsan and Bean [2003]
also observed a positive correlation between $p$ and heat-flow. 
The exponent $p$ was computed through the
estimation of the power-spectrum of the time-series of events. Their
figure 10 shows error bars on both $p$ and heat flow. Despite
the rather strong apparent correlation, we would like to stress 
that such results are difficult to interpret because
the data points are concentrated around average heat-flow values, and 
there are only a few data for very large
or very small heat-flow values. Thus, the dispersion of data is highly inhomogeneous around the mean
and such a biased sampling in heat flow may yield misleading results. Here, again
we propose to reconsider such an analysis with a more uniform sampling in heat flow and
according to the magnitude of main events.
Indeed, since our model predicts
that the magnitude and temperature effects are entangled, a careful
analysis of the magnitude relationship $p(M)$ is needed before testing
for a temperature or heat-flow effect.

We predict a negative correlation between $p$
and temperature which seems at first 
rather puzzling, as it predicts that the seismicity rate will relax more slowly at
larger temperatures (every other parameter being kept equal). 
This paradox is resolved by distinguishing between absolute seismicity rate
and relative decay rates. Indeed, expressions (\ref{mvmmcs}) and (\ref{gmjmwels})
show that the seismic rate can be decomposed into the product of two 
exponential terms
\be
\lambda({\vec r}, t) \sim e^{-\beta E_0({\vec r})}~e^{+\beta V \Sigma({\vec r}, t)}~.
\ee
The first exponential $e^{-\beta E_0({\vec r})}$
controls the absolute seismicity level and exhibits the
usual effect that, the higher the temperature $kT =1/\beta$, the larger
is the seismicity rate. The second exponential $e^{+\beta V \Sigma({\vec r}, t)}$
gives the dependence of the seismicity rate as a function of time due to 
stress interactions, as described in (\ref{mgmmwsls}) and following equations. It exhibits 
an inverse temperature effect: the larger the temperature, the smaller it is;
together with the dependence $\Sigma({\vec r}, t) \propto \ln (T/t)$, it is
responsible for the paradoxical temperature dependence of the $p$-value.
In summary, the larger the temperature, the larger is the absolute seismicity level
but the smaller is the $p$-value of Omori's law. More generally, distinguishing between absolute
and relative levels is essential when dealing with scale-invariant laws. The same
paradoxical effect occurs for instance with fractal dimensions: a large fractal dimension
does not necessarily imply a large density, since the density is the product of 
an absolute term setting the units and a scale-dependent part function of the 
fractal dimension: the fractal dimension only describes the relative values
of densities at two different scales, not their absolute values. Confusing these
two contributions to the density
is often done in the geophysical literature in which the dimension is 
incorrectly considered as a first-order measure of density; in reality, it is
only a relative measure comparing densities at different scales.
To come back to Kisslinger and Jones [1991]'s observations, rather than
large $p$-values associated with large heat flows, we suggest a more complicated
dependence involving a higher absolute seismicity level and a weaker dependence of $p$
on the mainshock magnitude as the temperature (heat flow) increases.
Everything being kept equal, as the temperature 
increases, the $p(M)$ values are predicted to decrease. This may perhaps also explain the 
careful laboratory observations of Carreker [1950]
who showed for Pt (platinium) that the strain rate exponent in creep experiments in the primary regime
decreases with temperature. Some other complications can occur if the integral
time scale $T$ also depends on the temperature, but if it is very large as we expect (at
least a few years),
its variation with the temperature $1/\beta$ will not have significant consequences
for the seismic aftershock decay rates studied here.

The decrease of $p$ with increasing temperature has been also documented
in a numerical simulation of the sandpile model of Christensen and Olami [1992]
in which elements break by static fatigue according to the rate (\ref{mvmmcs})
with (\ref{gmjmwels}). It was found [{\it Helmstetter}, 2002] that this decrease of $p$ with $T$
is the opposite of the prediction of the standard thermal activation model
when neglecting interactions between rupture and thus results fundamentally
from multiple interactions between events.

\subsubsection{Magnitude-dependence of the $p$-values in other works}

The increase of the $p$-value with the mainshock magnitude implies that 
aftershock sequences of large events 
decay at a faster rate than aftershock sequences of small events. Yet, they
have a much larger number of events and can thus be observed in general 
over longer times. According to our analysis, mainshocks of
magnitudes going from $5$ and $7.5$ (for which
the Omori law is generally inverted), the average
$p$-value increases from $0.9$ to $1.2$. These are the typical values generally
given in the literature. Our detection of a systematic increase of $p(M_L)$
was only made clear by extending the range of magnitudes over which the 
Omori law is tested.

There has been only few attempts to try to correlate $p$-values of individual aftershock time-series
with the magnitude of the mainshock (see [{\it Kisslinger and Jones}, 1991], for example,
in which no correlation was found which is not surprising as our analysis shows that
one has to stack many sequences following similar magnitude shocks in order to average
fluctuations out). The only other work we are aware, which uses stacked sequences, 
is the one of Bohnenstiehl et al. [2003].
These authors studied the time-clustering of events of magnitude $M > 3$ along the
Mid-Atlantic Ridge, using catalogs derived from the detection of T-waves. Each event is
quantified using a source level ($SL$), expressed in $dB$ units, which is a logarithmic
measure of its size. The earthquake catalog is then represented as a series of point
process events located at each earthquake's occurrence time, whose power spectrum can be
computed and fitted with a power law of exponent $\alpha_{ps}$ (they indeed use an Allan
factor analysis to determine $\alpha_{ps}$). Tuning the detection threshold $SL_0$ of
events, they note a tendency for $\alpha_{ps}$ to decrease as $SL_0$ increases. Since
the $p$-value is related to $\alpha_{ps}$ by $p=1-\alpha_{ps}$, 
then their observation confirms that $p$ decreases with
$M$ (since, due to the Gutenberg-Richter law, their power-spectrum is dominated by the statistics
related to the smallest magnitude events in the catalogue).

Mines offer meso-scale crustal laboratories to study superficial earthquakes. The sizes of
events are most often of magnitude $3$ and lower. As stated above, such events 
trigger a few aftershocks so that the recovery of an Omori law requires stacking
many events. Marsan et al. [1999] studied the time clustering of events
in the Creighton mine (Ontario, Canada), and built a stacked time-series which is
equivalent to ours if we consider that the size
of the aftershock area after each event is the size of the mine itself. 
Despite the fact that they did not remove
shocks already tagged as aftershocks from the main events list (which is a minor departure
from our procedure considering the small size of all events), they measured $p=0.4$.
They do not mention the magnitude range of events they used in their paper, but such a $p$-value
suggests (from our observations) that the majority of their shocks would roughly be 
of magnitude $1-3$, which is a very reasonable prediction for such events. According to our model, 
we should also take into account the difference of temperature to get a 
reliable quantitative comparison, but 
the low $p$-value found by Marsan et al. [1999] is already
in qualitative agreement with our model.

\subsubsection{Dependence of the $p$-value on the shear and normal
stress components}

The multifractal stress activation model may rationalize the empirical
observations of a dependence of the $p$-value on the shear and normal
stress components, which suggests the relevance of fluid pressure
[{\it Scholz}, 2002]. These properties which
cannot be accounted for by the Dieterich model [{\it Dieterich}, 1994]
can naturally arise from the impact of fluid pression on the stress redistribution.

\subsubsection{Distribution of seismic rates and of stress source strengths}

Our multifractal stress activation model can be further falsified
by comparing its prediction of the distribution 
of seismic rates ${\rm Pr}(\lambda)$, 
once the distribution (\ref{vcmnfdlqa}) of stress source strengths and the 
stress relaxation memory kernel (\ref{f,ma,fa}) are specified. As  
${\rm Pr}(\lambda)$ seems to be a power law with exponent $\mu \approx 1.5$
[Work in progress],
this seems to imply a truncation of the distribution of the stress source strengths.
More work is required to clarify this issue and will be reported elsewhere.

\subsubsection{Implications for prediction}

If true, our discovered multifractal Omori law has probably many important implications
for understanding earthquake patterns and for prediction, that need to be investigated
in details. The interplay between magnitude and decay rate found here leads to 
new interpretations of spatio-temporal patterns of seismicity. In particular, this 
will shed light on the underlying basis of
various pattern recognition techniques that tend to sort earthquakes
in terms of their magnitudes: according to our theory, different magnitude
classes which are controlled by the same underlying physics give rise
to distinct triggering signatures. 

\subsubsection{Lower bounds for a minimum earthquake magnitude}

The reported dependences (\ref{gnmgd}) and (\ref{gmmgsas}) of $p(M_L)$ 
suggests the existence of a lower bound $M_{\rm min}$ for the minimum magnitude of earthquake
able to trigger other events. Indeed, from the condition $p \geq 0$, we obtain
$M_{\rm min}> -2.3$ using (\ref{gnmgd}) and $M_{\rm min}> -3.7$ (\ref{gmmgsas}). The 
real uncertainty is difficult to estimate and is probably of the order of the difference
between these two values. Since there is no way we can address all known and 
unknown systematic error terms and uncertainties, the best way to estimate 
the uncertainty is by comparison of two different procedures with several different
implementation, as we have done. This leads to the estimate $M_{\rm min}> -3 \pm 1$.
Note that the existence of $M_{\rm min}$ does not imply that there are no 
earthquakes of smaller magnitudes, only that those smaller events do not play a
role in the triggering process (they do not trigger other events).

\acknowledgments
We are grateful to K. Felzer, Y.Y. Kagan, J.-F. Muzy and J. Vidale for stimulating discussions.
This work was partially supported by NSF-EAR02-30429 and by
the Southern California Earthquake Center (SCEC).



\end{article}

\clearpage

\clearpage

\clearpage

\clearpage

\clearpage

\clearpage

\clearpage

\clearpage

\clearpage

\clearpage

\clearpage

\clearpage

\clearpage

\clearpage

\clearpage

\clearpage

\end{document}